\documentclass[12pt]{article}

\usepackage{graphics}
\usepackage{graphicx}

\usepackage{epsfig}

\usepackage{natbib}

\usepackage{amssymb}
\usepackage{amsthm}
\usepackage{bm}
\usepackage{amsmath}
\usepackage{color}

\usepackage{geometry} 
\begin{document}

\title{Electro-mechanically guided growth and patterns}

\author{Y. Du$^{a,e}$, Y. Su$^{a,f}$, C. L\"{u}$^{b,c,d}$, W. Chen$^{a,c,d}$, M. Destrade$^{e,a}$\\[12pt]
$^a$Department of Engineering Mechanics, \\ Zhejiang University, Hangzhou 310027, P. R. China;\\[6pt]
$^b$Department of Civil Engineering,\\ Zhejiang University,  Hangzhou 310058, P.R. China; \\[6pt]
$^c$Key Lab of Soft Machines and Smart Devices of Zhejiang Province, \\ Zhejiang University, Hangzhou 310027, P.R. China; \\[6pt]
$^d$Soft Matter Research Center, \\ Zhejiang University, Hangzhou 310027, P. R. China; \\[6pt]
$^e$School of Mathematics, Statistics and Applied Mathematics,\\ NUI Galway, University Road, Galway, Ireland;\\[6pt]
$^f$Sonny Astani Department of Civil and Environmental Engineering,\\ University of Southern California, Los Angeles, CA 90089, USA}

\date{}

\maketitle


\begin{abstract}

Several experiments have demonstrated the existence of an elec\-tro-me\-cha\-ni\-cal effect in many biological tissues and hydrogels, and its actual influence on growth, migration, and pattern formation. 
Here, to model these interactions and capture some growth phenomena found in Nature, we extend volume growth theory to account for an electro-elasticity coupling.
Based on the multiplicative decomposition, we present a general analysis of isotropic growth and pattern formation of electro-elastic solids under external mechanical and electrical fields.
As an example, we treat the case of a tubular structure to illustrate an electro-mechanically guided growth affected by axial strain and radial voltage.
Our numerical results show that a high voltage can enhance the non-uniformity of the residual stress distribution and induce  extensional buckling, while a low voltage can delay the onset of wrinkling shapes and can also generate more complex morphologies.
Within a controllable range, axial tensile stretching shows the ability to stabilise the tube and help form more complex 3D patterns, while compressive stretching promotes instability.
Both the applied voltage and external axial strain have a significant impact on guiding growth and pattern formation.
Our modelling provides a basic tool for analysing the growth of electro-elastic materials, which can be useful for designing a pattern prescription strategy or growth self-assembly in Engineering.

\end{abstract}




\section{Introduction}



Growth and remodelling count among the most basic and essential biological activities, as they not only promote biodiversity but also ensure normal biological function and regeneration.
Internal genetic information, chemical stimuli, and physical conditions have been proved to affect the growth process, from the level of molecules and cells all the way to tissues and organs \citep{goriely2017mathematics, lewis2008signals, martin1998skeletal, zhao2009electrical}. 
At the level of tissues and organs, physical factors, including mechanical and electrical conditions, have shown strong impact on  residual stress accumulation and pattern evolution \citep{mendoncca2003directly, levin2014molecular, ahn2009relevance}.



With  \textit{volume growth theory} \citep{rodriguez1994stress, amar2005growth},  residual stress accumulation and pattern evolution are well explained as a result of re-balancing the incompatible volume swelling/absorption and excessive residual stress \citep{li2011surface, ciarletta2014pattern, balbi2015morphoelastic}. 
Moreover, as  residual stress is created and conserved throughout the whole growth process,  stress- or strain-dependent growth models have been proposed theoretically and verified experimentally \citep{Fung1991, du2018modified, du2019prescribing, du2019influence}.

\begin{figure}[h]
\begin{center}
\includegraphics[width=0.8\textwidth]{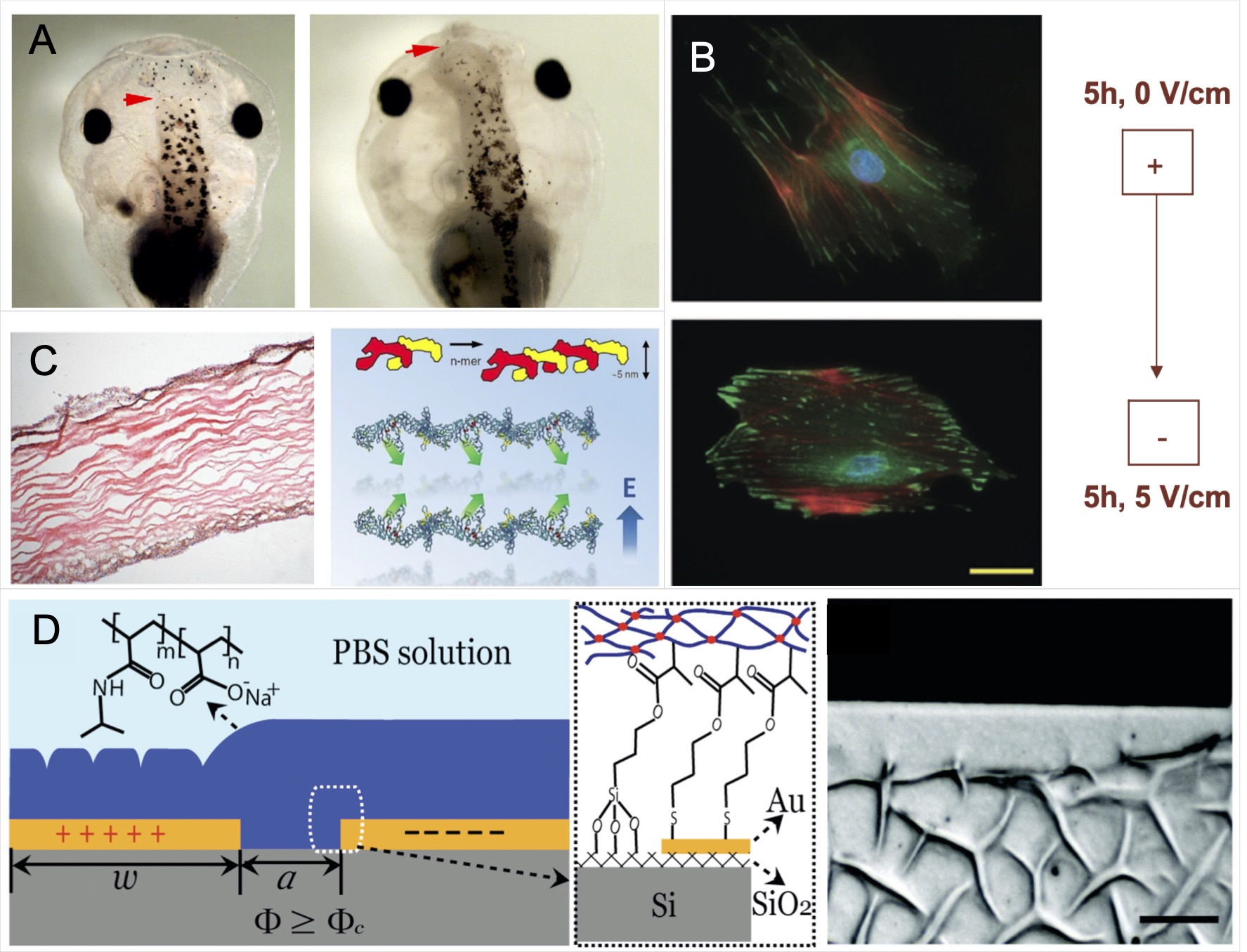}
\caption{(A). The electric field caused by mis-expression of ion channel during embryogenesis can make coherent changes in pattern, leading  (left) a normal forebrain to being (right) drastically increased in size \citep{levin2009bioelectric}; (B). Rat calvaria osteoblasts and fibroblasts subjected to the external electric field exhibit the ability to reorient and elongate cells in a perpendicular direction \citep{funk2009electromagnetic}; (C). The fibrous structure of elastin and schematics of the heat-tail configuration of tropoelastin monomers in which the envisioned molecular structure with a dipole moment that can be rotated by the applied electric field \citep{yang2011improving}; (D) Illustration of a device  triggering self-assembled monolayer anionic PNIPAM copolymer hydrogel by electric potentials from the underlying electrodes \citep{xu2013low}.}
\label{S1_fig9}
\end{center}
\end{figure}

However, it is worth recalling  that the growth and remodelling processes involve many complex physiochemical reactions.
In addition to mechanical factors, other physical fields such as an \textit{electric field} could also regulate the growth process.
Numerous experiments and protocols have indeed verified the practical impact of electric fields on tissue development and wound healing \citep{marino1970piezoelectric, jaffe1977electrical, jaffe1984electric}.
For example Figure \ref{S1_fig9}A shows that the electric field caused by mis-expression of ion channels during embryogenesis can produce coherent changes in patterns \citep{levin2009bioelectric}.
Another evidence (Figure \ref{S1_fig9}B) is that  rat calvaria osteoblasts and fibroblasts subjected to an external electric field exhibit the ability to reorient and elongate cells in a perpendicular direction \citep{funk2009electromagnetic}.

From the perspective of material properties, \cite{Fukada1957}  proved experimentally that bone exhibits the piezoelectric effect. 
Soon after, teeth, skin, nerve tissues, blood vessels, and dried collagen were also confirmed to exhibit an electro-mechanical  coupling effect \citep{anderson1968electrical, athenstaedt1970permanent, fukada1969piezoelectric, chae2018review}.
In addition, using  X-ray micro-diffraction experiments on hydroxyapatite unit cells of the bone, \citet{wieland2015investigation} revealed that the inverse piezoelectric effect could induce adequate strain levels to trigger a mechanism for bone growth.
Furthermore, by using  piezoelectric response force microscopy and molecular dynamics simulations \citep{liu2014ferroelectric, liu2012biological}, as shown in Figure \ref{S1_fig9}C,  the microscopic essence of the electro-mechanical effect of most soft bio-tissues was revealed to be due to the polar structure of tropoelastin. 
\cite{zelisko2015mechanism} established that the piezoelectric coefficient of tropoelastin is about 96.6 pC N$^{-1}$, which is a   remarkable value compared with other piezoelectric polymers such as the PVDF polymer, for which it is 33 pC N$^{-1}$.

On the other hand, many polymeric materials, such as VHB 4910, display a strong ability to imbibe  solvent and swell \citep{bosnjak2020experiments}.
In addition, \citet{kim2002electric} showed that the mechanical properties of some polyelectrolyte hydrogels and electro-active hydrogels, such as PVA/chitosan IPN, are related not only to their specific aqueous solutions but are also very sensitive to electrical stimuli.
In their experiments, a swollen polyelectrolyte hydrogel is placed between a pair of electrodes and bends in response to an applied electric field. 
The bending angle and the bending speed increase with an increase in the applied voltage and in the concentration of NaCl in the aqueous solution.

By taking advantage of this type of electro-mechanical response,  electrical stimuli have been used widely  to trigger self-assembled patterns of colour and fluorescence on demand \citep{wang2011electro, wang2011creasing, wang2012dynamic, wang2014cephalopod, bosnjak2020experiments}, and to design electrically-assisted iono-printing electro-active hydrogel actuation and drug delivery devices \citep{palleau2013reversible, agnihotri2005electrically, Choi2020}.
Figure \ref{S1_fig9}D shows the working principle of a device triggering self-assembled monolayer anionic PNIPAM copolymer hydrogels using electric potentials created by underlying electrodes \citep{xu2013low}. 
There, other swelling related factors, such as the temperature and ionic strength, are also considered.
The experimental results show that creases are  formed selectively above the anode and that patterns can be  controlled precisely through the electrode geometry.
However, the actual mechanism for electrically-driven crease formation on hydrogel surfaces is still not completely understood. 
Nonetheless, it is clear that the total stress, which includes the Maxwell stress and the mechanical stress, must be  coupled  to the swelling process.

Accordingly, for numerous examples of growing soft matter including bio-tissues and polymers, a comprehensive growth model including both electrical and mechanical fields is needed. 

By relying on nonlinear elasticity theory \citep{dorfmann2010nonlinear, dorfmann2019instabilities, su2018wrinkles, su2019finite}, we adopt an  electro-elastic free energy function to capture  large deformation and electro-mechanical coupling  during the growth process. 
According to the multiplicative decomposition method, the energy function of materials in the virtual configuration should  be assumed to be stress-free. 
Similar to  the treatment found in volume growth modelling, here the growth factor is included through the elastic deformation and the current residual stress and electric field can be obtained within the finite deformation regime. 
Finally, as the residual stress is affected by external mechanical and electrical fields, we also perform an incremental bifurcation analysis to show how these electro-mechanical loads affect  the generation  of patterns.

The paper is organised as follows. In Section 2, we present a general analysis of  isotropic growth and pattern formation for electro-elastic solids under external mechanical and electrical biasing fields.
In Section 3, we take a tubular electro-elastic solid as an example, and put the tube  under an axial stretch and a voltage in the radial direction.
Section 4 shows some numerical results for the growing neo-Hookean dielectric solid. 
In particular we investigate  the influence of axial stretch and external voltage on the growth and non-growth induced pattern generation.
In Section 5, we discuss the effects and significance of electro-mechanical growth and draw some conclusions.

\section{Governing equations}



\subsection{Finite growth with external electro-mechanical fields}


Consider a continuous electro-elastic solid growing from an undeformed stress-free configuration $\mathcal{B}_0$ and  subjected to external mechanical and electrical stimuli. 
It  reaches a grown and residually stressed configuration $\mathcal{B}$, where  the position vector  $\boldsymbol x$ corresponds to the position vector $\boldsymbol X$ in $\mathcal{B}_0$. 
The deformation gradient tensor is  $\boldsymbol{F}={\partial{\boldsymbol x}}/{\partial \boldsymbol X}$. 

\begin{figure}[htbp]
\begin{center}
\includegraphics[width=.7\textwidth]{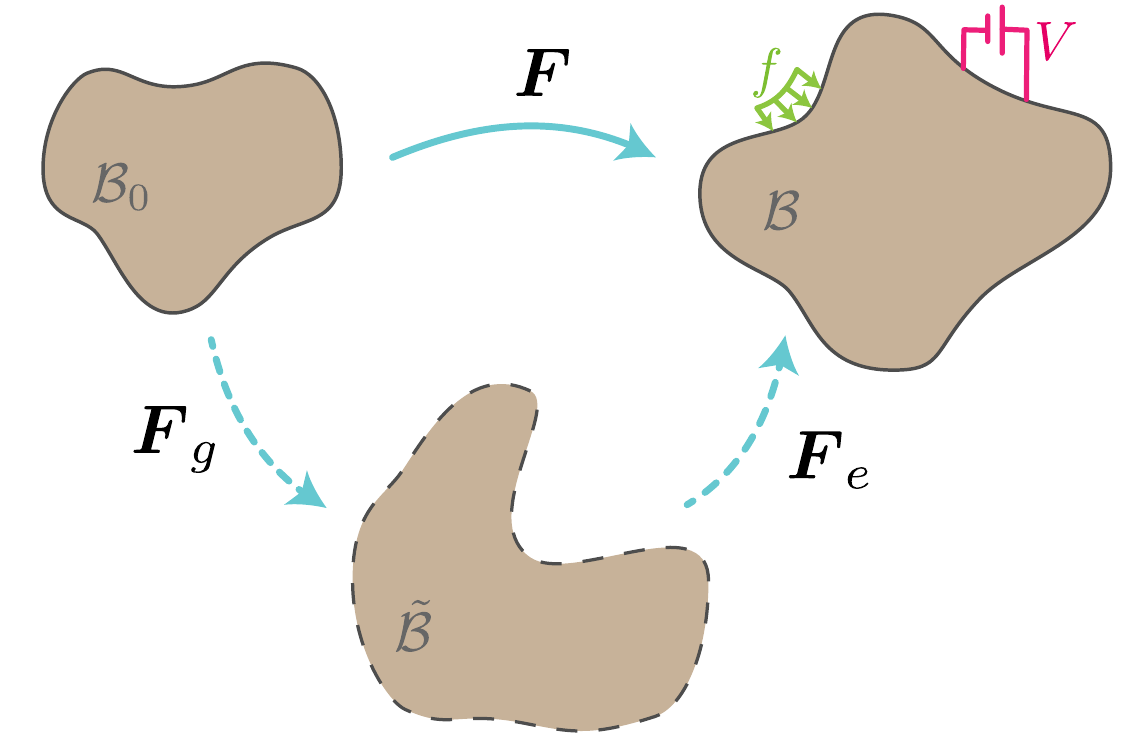}
\caption{The multiplicative decomposition of the volume growth modelling subject to the external traction and voltage.}
\label{Fig2}
\end{center}
\end{figure}

Using the multiplicative decomposition of volume growth modelling \citep{rodriguez1994stress}, see Figure \ref{Fig2}, we decompose the elastic deformation tensor $\boldsymbol{F}_e$ as 
\begin{equation}
\boldsymbol{F}_e=\boldsymbol{F} \boldsymbol{F}_g^{-1},
\label{eq1}
\end{equation}
where $\boldsymbol{F}_g$ is the pure growth deformation tensor, with $J_g=\det \boldsymbol{F}_g$ tracking its volume changes.
Note that, in the study, we hypothesise that the pure growth deformation is decoupled with  both applied electric field and external mechanical tractions.
Taking the material as incompressible, we impose that only isochoric elastic deformations are possible, so that 
\begin{equation}
J_e=\det \boldsymbol{F}_e=1,
\label{eq2}
\end{equation}
at all times.

According to nonlinear electro-elastic theory, we may track the impact of electro-mechanical stimuli on growth by taking a free energy density in the form $\Omega=\Omega(\boldsymbol{F}_e, \tilde{\boldsymbol{D}})$ in  the virtual stress-free configuration $\tilde{\mathcal{B}}$. 
Here, $\tilde{\boldsymbol{D}}$ is the electric displacement vector with respect to the virtual stress-free configuration, and its corresponding electric field is denoted $\tilde {\bm E}$.
In addition, with respect to the reference configuration, we call $\bm D_l $  the Lagrangian electric displacement and $\bm E_l$ the Lagrangian electric field vector. 
In the current configuration, we call $\bm D $ the current electric displacement  and $\bm E$ the electric field vector.

\color{black}
In the absence of free charges and currents, these electrical quantities  satisfy  
\begin{align}
\text{div} \, \boldsymbol{D=0}, \qquad \text{curl}\,\boldsymbol{E=0},\qquad \text{Div} \, \boldsymbol{D}_l= \boldsymbol 0, \qquad \text{Curl}\,\boldsymbol{E}_l  = \boldsymbol 0,
\end{align}
where curl and div are respectively the curl and divergence operators with respect to $\bm x$, and Curl and Div are respectively the curl and divergence operators with respect to $\bm X$.
Then, recall that \textit{Nanson's} formula  $\bm{n} da=J\bm{F^{-T} N} dA$ connects the current ($da$) and referential ($dA$) area elements, where $J=\det \bm F$ is the volume change with growth, and $\bm n$ and $\bm N$ are the outward unit vectors normal to surface elements in the current configuration and reference configurations, respectively.
Hence we have 
\begin{align}
\int_\mathcal{B} \text{div} \bm D \, dv&=\int_{\partial \mathcal{B}} \bm{D \cdot  n} \, da
=\int_{\partial \mathcal{B}_0} J \left(\bm{F^{-1}D\right) \cdot  N} dA \notag\\
&=\int_{\mathcal{B}_0} \text{Div} \left(J \bm{F^{-1}D} \right)dV = 
\int_{\mathcal{B}_0} \text{Div} \bm D_l \, dV=0,
\end{align}
and
\begin{align}
\int_{ \varphi} (\text{curl} \, \bm {E}) \bm{\cdot n}\, da
&=\int_{\partial \varphi}  \bm {E \cdot dx}
=\int_{\partial \varphi_0}   \left(\bm F^T \bm E\right) \bm{\cdot dX}\notag\\
&=\int_{ \varphi_0} \text{Curl}  \left(\bm F^T \bm E\right) \bm{\cdot N} dA
=\int_{ \varphi_0} (\text{Curl} \, \bm E_l) \bm{\cdot N} dA=0,
\end{align}
where $\varphi$  is an open surface in the current configuration and  $\partial \varphi$  is a closed curve bounding $\varphi$, defined in the usual sense relative to the unit normal $\bm n$ to $\varphi$, and  $\varphi_0$ and $\partial \varphi_0$ are their referential counterparts.
Hence, we have the connections 
\begin{align}
\boldsymbol {D}=J^{-1} \boldsymbol {F} \boldsymbol {D}_l , \qquad \boldsymbol{E} = \boldsymbol {F}^{-T} {\boldsymbol {E}_l},
\end{align}
Similarly, performing the same manipulations of the quantities in the virtual stress-free configuration, we obtain the following connections for these electrical fields,
\begin{align}
\boldsymbol {D}=J^{-1} \boldsymbol {F} \boldsymbol {D}_l = \boldsymbol {F}_e \tilde{\boldsymbol {D}}, \qquad \boldsymbol{E} = \boldsymbol {F}^{-T} {\boldsymbol {E}_l}=\boldsymbol {F}_e^{-T} \tilde {\bm E}.
\label{eq4}
\end{align}
\color{black}

Thus, for an incompressible solid with energy function \color{black} $\Omega(\boldsymbol{F}_e$, $\tilde{\boldsymbol{D}})-p(J_e-1)$, we obtain the stress and the electric field tensor with respect to the virtual stress-free configuration as \color{black} 
\begin{align}
{\tilde {\boldsymbol S}}=\frac{\partial \Omega}{\partial \boldsymbol{F}_e}-p\boldsymbol{F}_e^{-1}, 
\qquad 
{\tilde {\boldsymbol{E}}}=\frac{\partial \Omega}{\partial \tilde{\boldsymbol{D}}},
\label{eq3-1}
\end{align}
respectively, where $p$ is an arbitrary Lagrange multiplier, to be found from boundary and/or initial conditions.
Then, the nominal stress $\boldsymbol S$ and Lagrangian electric field $\boldsymbol{E}_l$ with respect to the reference configuration are \begin{align}
{\boldsymbol S}=J_g \bm F_g^{-1}\left(\frac{\partial \Omega}{\partial \boldsymbol{F}_e}-p\boldsymbol{F}_e^{-1} \right), 
\qquad 
{\boldsymbol{E}_l}= \bm F_g^T \frac{\partial \Omega}{\partial \tilde{\boldsymbol{D}}},
\label{eq3}
\end{align}
respectively, and the Cauchy stress ${\boldsymbol \sigma}$ and the current electric field ${\boldsymbol E}$ in the current configuration are, therefore,
\begin{align}
{\boldsymbol \sigma}=\color{black} \bm F_e \color{black} \frac{\partial \Omega}{\partial \boldsymbol{F}_e}-p\boldsymbol{I}, 
\qquad
{\boldsymbol{E}}= \bm F_e^{-T} \frac{\partial \Omega}{\partial \tilde{\boldsymbol{D}}},
\label{eq3}
\end{align}
respectively.

In the absence of  body forces, the equilibrium equation of the Cauchy stress reads
\begin{align}
\text{div} \, \boldsymbol{\sigma=0},
\label{eq6}
\end{align}
and the boundary conditions, in the absence of exterior electric fields, are
\begin{align}
\boldsymbol{\sigma^T n=t_a},\qquad \boldsymbol{D\cdot n}=q_e,\qquad \boldsymbol{E \times n=0},
\end{align}
where $\bm n$ is the outward unit  vector normal to  surface elements in the current configuration, $\bm{t_a}$ is the prescribed mechanical traction, and $q_e$ is the surface charge density on the boundary.

\color{black} According to the nonlinear electro-elastic theory developed by \cite{dorfmann2005nonlinear, dorfmann2010nonlinear,dorfmann2019instabilities}, \color{black} for isotropic, incompressible electro-elastic materials, the free energy function \color{black} $\Omega(\boldsymbol{F}_e$, $\tilde{\boldsymbol{D}})-p(J_e-1)$, \color{black} can  also be written as a function of the following  five invariants
\begin{align}
&I_1=\text{tr} \,\boldsymbol C_e,
&& I_2=\tfrac{1}{2}\left[ \left(\text{tr}\, \boldsymbol C_e\right)^2-\text{tr} \left(\boldsymbol C_e^2\right)\right], 
&& \notag \\
&I_4=\tilde{\boldsymbol D} \bm\cdot \tilde{\boldsymbol D},
&&I_5=\tilde{\boldsymbol D} \bm\cdot \boldsymbol C_e \tilde{\boldsymbol D},
&&I_6=\tilde{\boldsymbol D} \bm \cdot \boldsymbol C_e^2 \tilde{\boldsymbol D},
\label{eq8}
\end{align}
where $\boldsymbol C_e=\boldsymbol F_e^T \boldsymbol F_e$ is the right Cauchy-Green deformation tensor
(note that the third principal invariant $I_3=\det \boldsymbol C_e$ is equal to 1 at all times because of incompressibility \eqref{eq2}).
Then, according to Eq.~\eqref{eq4}, the Cauchy stress and the electrical field follow as
\begin{align}
  & \bm{\sigma}= 2 \Omega_1\bm B_e + 2 \Omega_2 (I_1 \bm B_e -\bm B_e^2)-p \bm I+2\Omega_5 \bm{D}\otimes\bm{D} +2\Omega_6(\bm{D}\otimes\bm{B}_e \bm D + \bm{B}_e \bm{D}\otimes\bm{D}),\notag \\
&\bm E=
 2( \Omega_4 \bm B_e^{-1}\bm{D}+\Omega_5 \bm{D}+\Omega_6 \bm{B}_e \bm{D}),
 \label{eq9}
\end{align}
where $\bm B_e=\boldsymbol F_e \boldsymbol F_e^T$  is the left Cauchy-Green deformation tensor and $\Omega_i = {\partial \Omega}/{\partial I_i}$.
It follows that because $\bm B_e$ is related to the growth deformation $\bm F_g$ and because $\bm D$ is determined by the electric field $\bm E$ and $\bm B_e$, the residual stress must depend both on  the growth factor and on external mechanical and electric fields. 
And so must  the growth-induced patterns. 


\subsection{Instability analysis}


To figure out the effects of  external mechanical and electric fields on  growth-induced pattern evolution, we rely on an incremental theory to analyse stability after growth.

First, we superimpose an infinitesimal incremental displacement $\bm{\dot{x}} = \bm{\dot\chi}(\bm X)$ on the current configuration $\mathcal{B}$ with respect to the reference configuration $\mathcal{B}_0$ and an incremental electric displacement $\dot{\tilde{\bm D}}$ with respect to the virtual stress-free configuration $\tilde {\mathcal{B}}$. 
The incremental displacement gradient tensor with respect to the reference configuration $\mathcal{B}_0$ is then   $\bm{\dot {F}}={\partial \bm{\dot{\chi}}}/{\partial \bm X}$, and with respect to the current configuration $\mathcal{B}$ it is $\bm{\dot{F}}_I={\partial \bm{\dot{\chi}}}/{\partial \bm x}$. 
Hence, we have the connection
\begin{equation}
\bm{\dot{F}}= \bm{\dot{F}}_I \bm F.
\end{equation}
Recall that the growth deformation $\bm F_g$ is independent of the elastic deformation. 
The incremental displacement and electric displacement are infinitesimal and independent of the growth deformation, so that we  also have 
\begin{equation}
\bm{\dot{F}}_e= \bm{\dot{F}}_I \bm F_e,
\label{eq38}
\end{equation}
where $\bm{\dot{F}}_e$ is the increment of the purely elastic deformation tensor $\bm F_e$.
Further, we find that the incremental incompressibility condition reads
\begin{equation}
\text{tr} \,\dot{\bm F}_I  =\bm 0.
\label{eq14-1}
\end{equation}

We now linearise the expressions for the stress measures. 
We obtain the incremental nominal stress and incremental  Lagrangian electric field as 
\begin{align}
&  \boldsymbol{\dot{S}} = J_g \bm F_g^{-1}\left(\mathcal{\bm A}_e \bm{\dot{F}}_e +\bm \Gamma \dot{\tilde{\bm D}} - \dot p \bm F_e^{-1} + p \bm F_e^{-1} \bm{\dot{F}}_e \bm F_e^{-1}\right),\notag \\ 
& \boldsymbol{\dot{E}}_l = \bm F_g^T \left( \bm \Gamma \bm{\dot{F}}_e  + \bm{\mathcal{K}} \dot{\tilde{\bm{{D}}} }\right),
\end{align}
where $\dot p$ is the increment of the Lagrange multiplier, and $\bm{\mathcal{A}}_e$, $\bm \Gamma$, and $\bm{\mathcal{K}}$ are respectively, fourth-, third- and second-order tensors, the \textit{electro-elastic moduli tensors}.
Their components are  \citep{dorfmann2010nonlinear}
\begin{equation}
{\mathcal{A}}_{e\alpha i\beta j}=\frac{\partial^2 \Omega}{\partial F_{ei\alpha} \partial F_{ej\beta}}, \quad
 \Gamma_{\alpha i \beta}=\frac{\partial^2 \Omega}{\partial F_{ei\alpha} \partial \tilde D_{\beta}}, \quad
{\mathcal{K}}_{\alpha \beta}=\frac{\partial^2 \Omega}{\partial \tilde D_{\alpha} \partial \tilde D_{\beta}}.
\end{equation}

Then, using   Nanson's formulas \eqref{eq4} and Eq.~\eqref{eq38}, we  further obtain the incremental nominal stress and Lagrangian electric field in their push-forward (or updated) form  as
\begin{align}
& \boldsymbol{\dot{S}}_I = J^{-1} \bm F \boldsymbol{\dot{S}}
= \mathcal{\bm A}_I  \bm{\dot{F}}_I + \bm \Gamma_I  \dot{\tilde {\bm D}}_I - \dot p \bm I + p \bm{\dot{F}}_I,
\notag \\ 
& \boldsymbol{\dot{E}}_{lI} = \bm F^{-T} \boldsymbol{\dot{E}}_l =
 \bm \Gamma_I \bm{\dot{F}}_I + \bm{\mathcal{K}}_I \dot{\tilde {\bm D}}_I.
 \label{eq16-1}
 \end{align}
Here $\dot{\tilde {\bm D}}_I = \bm F_e \dot{\tilde {\bm D}}$, and  $\bm{\mathcal{A}}_I$, $\bm \Gamma_I$, and $\bm{\mathcal{K}}_I$ are the updated electro-elastic moduli tensors, with  components 
\begin{equation}
{\mathcal{A}}_{Ipiqj}=F_{ep\alpha} F_{eq \beta} {\mathcal{A}}_{e\alpha i\beta j},\quad
\Gamma_{Ip i q}= F_{ep \alpha} F_{e\beta q}^{-1}   \Gamma_{\alpha i \beta},\quad
{\mathcal{K}}_{Ip q}=F_{e\alpha p }^{-1 } F_{e\beta q}^{-1}  {\mathcal{K}}_{\alpha \beta},
\end{equation}
and symmetries
\begin{equation}
{\mathcal{A}}_{Ipiqj}={\mathcal{A}}_{Iqjpi},
\qquad
\Gamma_{Ip i q}=\Gamma_{Ii p q},
\qquad
{\mathcal{K}}_{Ip q}={\mathcal{K}}_{I qp}.
\end{equation}
Moreover, we note, using the incremental form of the symmetry condition of the Cauchy stress $\bm F \bm S=\bm {\left(FS\right)^T}$, that the following connections apply,
\begin{equation}
{\mathcal{A}}_{epiqj}^I-{\mathcal{A}}_{eipqj}^I=\left(\sigma_{pq}+p\delta_{pq}\right)\delta_{ij}-\left(\sigma_{iq}+p\delta_{iq}\right)\delta_{pj}.
\label{eq20-1}
\end{equation}

Finally, the incremental equilibrium equations read
\begin{equation}
\text{div}\, \dot{ \bm{S}_I}=\bm 0,
\qquad
\text{div}\, \dot{\tilde {\bm D}}_I=\bm 0,
\qquad
\text{curl}\, \dot{\bm E}_{lI}=\bm 0.
\end{equation}

Now recall that the  solid is assumed  to grow in the absence of body forces, free charges and currents, and that the increments of electrical variables in the surrounding vacuum are disregarded.
 Hence, the incremental form of the electric and mechanical boundary conditions read
\begin{equation}
\dot{\bm{S}}_I^T \bm n=\dot {\bm t}_{Ia},
\qquad 
\dot{\tilde {\bm D}}_I \bm{\cdot  n} = \dot q_e,
\qquad
\dot{ \bm E}_{lI} \times \bm n=\bm 0,
\end{equation}
where $\dot {\bm t}_{Ia}$ and $\dot q_e$ are the incremental mechanical traction and surface charge density per surface element of the boundary $\partial \mathcal{B}$. 

Our goal is to find nontrivial solutions that mathematically satisfy the incremental equilibrium equations and boundary conditions,  indicating critical states. 
Once the critical value of the initial instability is found, we can then obtain the corresponding morphology of this electro-mechanically controlled growth.


\section{Growing tube under external electro-mechanical fields}



\subsection{Residual stress and electric field after growth}


Here we take a growing tubular electro-elastic solid as an example.
The tube is under an axial stretch and a voltage is applied in the radial direction.
It grows isotropically and independently of these external electro-mechanical loads. 
\begin{figure}[htbp]
\begin{center}
\includegraphics[width=.7\textwidth]{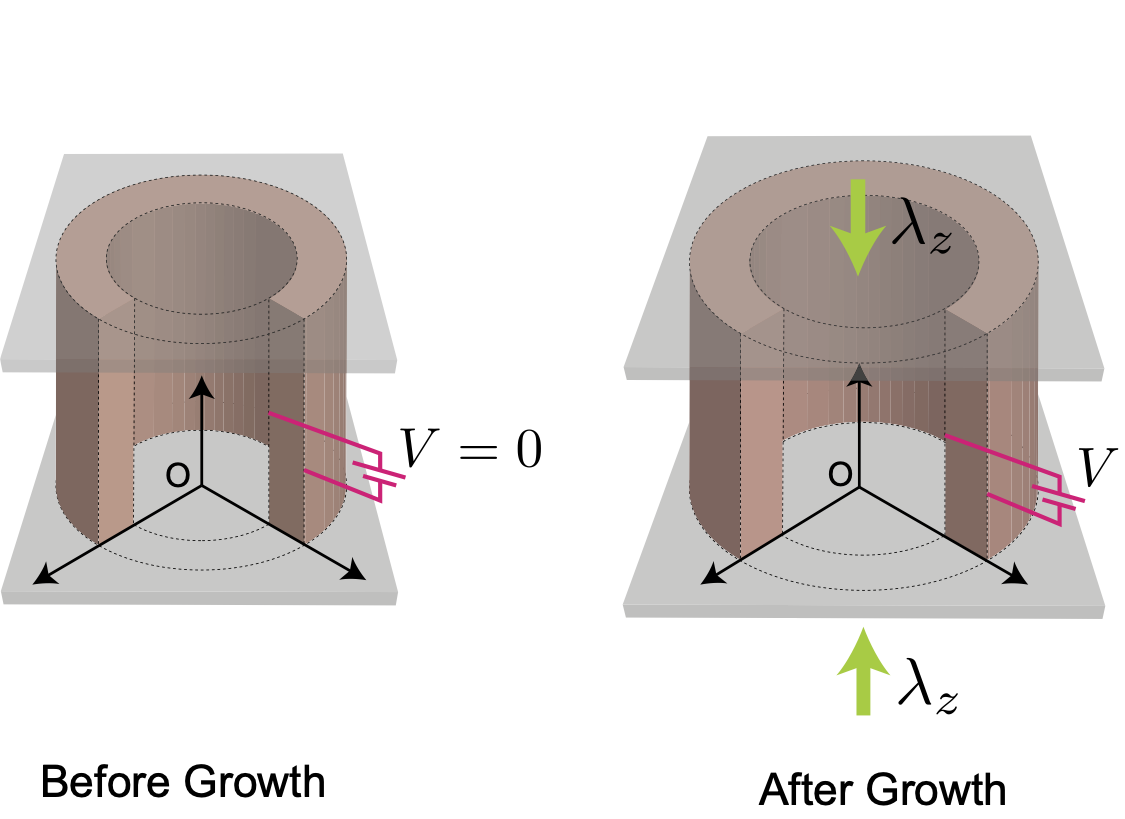}
\caption{The growing tubular structure made of an electro-elastic solid, where the axial stretch $\lambda_z$ is constrained and the voltage $V$ is applied across  the radial direction.}
\label{Fig3}
\end{center}
\end{figure}

As shown in Figure \ref{Fig3}, before growth, the tube is located in the region 
\begin{align}
  R_i \leq R \leq R_o,~ -\pi \leq \Theta \leq \pi,~ 0 \leq Z \leq L,
\end{align}
which is the reference configuration. 
In the current configuration, it occupies the region  
\begin{align}
  r_i \leq r \leq r_o,~ -\pi \leq \theta \leq \pi,~ 0 \leq z \leq \ell.
\end{align}
Then, using Eq.~\eqref{eq1}, we obtain the deformation gradient tensors as
\begin{align}
 \bm F=\left[
 \begin{matrix}
 \dfrac{\partial r}{\partial R} &0 &0\\
 0 &\lambda &0\\
 0 &0 &\lambda_z\\
 \end{matrix}\right], \qquad
  \bm F_g=\left[
 \begin{matrix}
g &0 &0\\
 0 &g &0\\
 0 &0 &g\\
 \end{matrix}\right],\qquad
   \bm F_e=\left[
 \begin{matrix}
g^{-1} \dfrac{\partial r}{\partial R} &0 &0\\
 0 &g^{-1}\lambda &0\\
 0 &0 &g^{-1}\lambda_z\\
 \end{matrix}\right],
 \label{eq23-1}
\end{align}
where $\lambda=r/R$, $\lambda_z=\ell/L$. 
Enforcing the incompressibility condition, we have
\begin{equation}
\frac{\partial r}{\partial R}=g^3\lambda^{-1}\lambda_z^{-1},
\label{eq14}
\end{equation}
which further gives the geometrical connection
\begin{equation}
R^2-R_i^2=g^{-3}\lambda_z(r^2-r_i^2).
\label{eq15}
\end{equation}

As shown in Figure \ref{Fig3}, the axial strain is applied externally, and the voltage is applied along the radial direction. 
Thus, the nominal electric field and displacement in the reference configuration are of the form
\begin{equation}
{\bm E_l}=\left[E_R,~0,~0\right]^T,\qquad
{\bm D_l}=\left[D_R,~0,~0\right]^T.
\label{eq26-1}
\end{equation}
Recalling the connections in Eq.~\eqref{eq4}, the electric field and the displacement with respect to the virtual configuration are
\begin{align}
&\tilde {\bm E}=\left[\tilde E_r,~0,~0\right]^T=\left[g^{-1} E_R,~0,~0\right]^T,\\
&\tilde{\bm D}=\left[\tilde D_r,~0,~0\right]^T=\left[g^{-2} D_R,~0,~0\right]^T.
\end{align}
In addition, the true electric field and the displacement in the current configuration are
\begin{align}
& \bm E=\left[E_r,~0,~0\right]^T=\left[g^{-3} \lambda\lambda_z E_R,~0,~0\right]^T,\notag \\
& \bm D=\left[D_r,~0,~0\right]^T=\left[\lambda^{-1}\lambda_z^{-1} D_R,~0,~0\right]^T.
\end{align}
Then the invariants in Eq.~\eqref{eq8} reduce to
\begin{align}
&I_1=g^4\lambda^{-2}\lambda_z^{-2}+g^{-2}\lambda^2+g^{-2}\lambda_z^2, 
\qquad I_2=g^{-4}\lambda^{2}\lambda_z^{2}+g^{2}\lambda^{-2}+g^{2}\lambda_z^{-2},\notag \\
&I_4=g^{-4} D_R^2,
\qquad I_5= \lambda^{-2}\lambda_z^{-2}  D_R^2,
\qquad I_6= g^4 \lambda^{-4}\lambda_z^{-4} D_R^2.
\label{eq18}
\end{align}

Now using Eqs.~\eqref{eq9} and \eqref{eq18}, we find the following non-zero components of the Cauchy stress  tensor $\bm \sigma$ and of the current electric field vector $\bm E$:
\begin{align}
 & {\sigma_{rr}}= 2 \Omega_1 g^4 \lambda^{-2} \lambda_z^{-2}  + 2 \Omega_2 g^2(\lambda^{-2}+\lambda_z^{-2})-p
+2\Omega_5 \lambda^{-2} \lambda_z^{-2} D_R^2 +4\Omega_6 g^4 \lambda^{-4} \lambda_z^{-4} D_R^2,\notag \\
&  \sigma_{\theta\theta}= 2 \Omega_1 g^{-2} \lambda^{2}  + 2 \Omega_2 (g^2\lambda_z^{-2}+g^{-4}\lambda^{2}\lambda_z^{2})-p,\notag\\
&    \sigma_{zz}= 2 \Omega_1 g^{-2} \lambda_z^{2}  + 2 \Omega_2 (g^2\lambda^{-2}+g^{-4}\lambda^{2}\lambda_z^{2})-p,\notag\\
 &  E_r=
 2( \Omega_4 g^{-4} \lambda \lambda_z +\Omega_5 \lambda^{-1} \lambda_z^{-1}+\Omega_6 g^4 \lambda^{-3} \lambda_z^{-3})D_R.
\end{align}

We may combine these expressions to obtain the following compact relations:
 \begin{align}
\sigma_{\theta\theta}-\sigma_{rr}=\lambda \dfrac{\partial \Omega}{\partial \lambda},
\qquad\sigma_{zz}-\sigma_{rr}=\lambda_z \dfrac{\partial \Omega}{\partial \lambda_z},
\qquad E_r=\lambda \lambda_z \dfrac{\partial \Omega}{\partial D_R}.
\label{eq21}
\end{align}
Then the equilibrium equation in Eq.~\eqref{eq6}$_1$ can  be rewritten as
\begin{align}
\frac{\partial \sigma_{rr}}{\partial r}=\frac{\lambda}{r} \frac{\partial \Omega}{\partial \lambda}.
\end{align}
Furthermore, noticing that $\frac{d r}{d \lambda}=\frac{g^3 r}{\lambda(g^3-\lambda_z \lambda^2)}$,
the Cauchy stress at any position $r$ is obtained by integration as
\begin{equation}
\sigma_{rr}(r)=\int_{r_i}^{r} \frac{\lambda}{r} \frac{\partial \Omega}{\partial \lambda}dr- \sigma_{rr}(r_i)=\int_{\lambda_i}^{\lambda} \frac{\partial \Omega}{\partial \lambda} \frac{g^3}{g^3-\lambda_z \lambda^2}d\lambda- \sigma_{rr}(\lambda_i),
\label{eq23}
\end{equation}
where $\lambda_i=r_i/R_i$.
Assuming that the inner face and outer surface are both free of mechanical traction, so that 
\begin{align}
 \sigma_{rr}(r_i)=0,
 \qquad \sigma_{rr}(r_o)=0,
\end{align}
we arrive at 
\begin{equation}
\sigma_{rr}(\lambda)=\int_{\lambda_i}^{\lambda} \frac{\partial \Omega}{\partial \lambda} \frac{g^3}{g^3-\lambda_z \lambda^2}d\lambda,
\qquad 
0=\int_{\lambda_i}^{\lambda_o} \frac{\partial \Omega}{\partial \lambda} \frac{g^3}{g^3-\lambda_z \lambda^2}d\lambda,
\label{eq25}
\end{equation}
where $\lambda_o=r_o/R_o$.
According to  Eq.~\eqref{eq21}, the circumferential stress $\sigma_{\theta\theta}$ and the axial stress $\sigma_{zz}$ can then  be obtained from the radial stress $\sigma_{rr}$. 

Turing now to Maxwell's equation in Eq.~\eqref{eq6}$_2$, we see that the equilibrium equation for the current electric field reduces to
\begin{equation}
\frac{1}{r}\frac{\partial (rD_r)}{\partial r}=0,
\end{equation}
so that  
\begin{equation}
D_r=\frac{c}{r},
\end{equation}
where $c$ is an integration constant, which can be determined by specifying the voltage in the current configuration.
As the electric field is the negative gradient of the the electric potential $\phi$: $\bm E=-\text{grad} \phi$, and the voltage $V$ is the potential difference $\phi_i-\phi_o$ between the inner and outer surfaces, we  then find 
\begin{equation}
V=\phi_i-\phi_o=\int_{r_i}^{r_o} E_r dr=\int_{r_i}^{r_o} \frac{D_r}{\varepsilon} dr=\frac{c}{\varepsilon}\ln{\frac{r_o}{r_i}},
\label{eq26}
\end{equation}
where $\bar r_o=r_o/r_i$ is a dimensionless measure of the outer radius. 
Here we assumed \textit{ideal electro-elasticity} for the calculation, that is, we assume that $E_r=D_r/\varepsilon$, where $\varepsilon$ is the dielectric permittivity.
Hence, with Eq.~\eqref{eq26}, the electric electric displacement and field  finally read
\begin{equation}
D_r=\frac{\varepsilon V}{r \ln\left( r_o/r_i\right)},
\qquad E_r=\frac{V}{r \ln \left( r_o/r_i\right)}.
\label{eq27}
\end{equation}

Then, within the framework of nonlinear electro-elasticity theory \citep{dorfmann2010nonlinear}, we split the free energy function of the electro-elastic material into
\begin{align}
\Omega(\bm F_e, \tilde {\bm D})&=W(\bm F_e)+ \Omega^* (\bm F_e, \tilde {\bm D}),
\end{align}
where $W(\bm F_e)$ is the elastic energy function part and  $\Omega^* (\bm F_e, \tilde {\bm D})$ is the electro-elastic energy function part. 
For the \textit{ideal electro-elastic solid} \color{black}\citep{dorfmann2005nonlinear, dorfmann2010nonlinear,dorfmann2019instabilities, zhao2007method}\color{black},  $\Omega^* (\bm F_e, \tilde {\bm D}) = I_5/(2\varepsilon)$, which we now adopt.

The Cauchy stress of  ideal electro-elastic solids then follows from Eq. \eqref{eq23} as
\begin{equation}
\sigma_{rr}(r) = \int_{\lambda_i}^{\lambda} \frac{\partial W}{\partial \lambda} \frac{g^3}{g^3-\lambda_z \lambda^2}d\lambda
 - \dfrac{1}{\varepsilon}\int_{\lambda_i}^{\lambda} {\lambda^{-3}\lambda_z^{-2} D_R^2} \frac{g^3}{g^3-\lambda_z \lambda^2}d\lambda.
\label{eq46-1}
\end{equation}

Then we introduce the following dimensionless quantities,
\begin{align}
&\bar\Omega=\Omega/\mu,
&& \bar W=W/\mu,
&& \bar r_o=r_o/r_i,
&&\bar R_o=R_o/R_i,\notag\\
&\bar \sigma_{rr}=\sigma_{rr}/\mu,~
&&\bar V=\frac{V}{R_o-R_i}\sqrt{\frac{\varepsilon}{\mu}},
&&\bar D_r=D_r/\sqrt{\varepsilon \mu}.
&&
\end{align}
where $\mu>0$ is the initial shear modulus in the absence of electric field. 
Using Eqs.~\eqref{eq25} , \eqref{eq27}, and \eqref{eq46-1},
we arrive at the following expressions for the dimensionless voltage,
\begin{equation}
\bar V=\frac{\bar r_o \lambda_i \ln \bar r_o}{\bar R_o-1}
\sqrt{\frac{2}{1-\bar r_o^2}  \int_{\lambda_o}^{\lambda_i} \frac{\partial \bar W}{\partial \lambda}\frac{g^3}{g^3-\lambda_z \lambda^2}d\lambda}.
\label{dimensionless-voltage}
\end{equation}
and the dimensionless Cauchy stress 
\begin{multline}
\bar \sigma_{rr}(\lambda)=
\frac{\lambda_o^2 \bar R_o^2(\lambda_i^2-\lambda^2)}{\lambda^2(\lambda_o^2 \bar R_o^2-\lambda_i^2)(g^{-3}\lambda_z \lambda_i^2-1)}  \int_{\lambda_o}^{\lambda_i} \frac{\partial \bar W}{\partial \lambda}\frac{g^3}{g^3-\lambda_z \lambda^2}d\lambda  
+\int_{\lambda_i}^{\lambda} \frac{\partial \bar W}{\partial \lambda}\frac{g^3}{g^3-\lambda_z \lambda^2}d\lambda.
\label{dimensionless-stress}
\end{multline}



\color{black}
\subsection{Incremental equations}


Now we superimpose an incremental elasto-electric perturbation on the deformed configuration to study the stability of the tube after growth. 
The components of these increments are written in cylindrical coordinates,
\begin{equation}
u_i=u_i\left(r, \theta, z \right), \qquad \dot{\tilde {D}}_{Ii}=\dot{\tilde {D}}_{Ii}\left(r, \theta, z \right).
\end{equation}
The incremental deformation gradient tensor $\dot {\bm {F}}_I$ follows as
\begin{equation}
\dot {\bm {F}}_I = \left[ \begin{matrix}
   \dfrac{\partial u_r}{\partial r} & \dfrac{1}{r}\left(\dfrac{\partial u_r}{\partial \theta}-u_\theta\right) & \dfrac{\partial u_r}{\partial z}  \\[10pt]
   \dfrac{\partial u_\theta}{\partial r} & \dfrac{1}{r}\left(\dfrac{\partial u_\theta}{\partial \theta}+u_r\right) & \dfrac{\partial u_\theta}{\partial z}  \\[10pt]
   \dfrac{\partial u_z}{\partial r} & \dfrac{1}{r}\dfrac{\partial u_z}{\partial \theta} & \dfrac{\partial u_z}{\partial z}  
\end{matrix} \right],
\label{eq52}
\end{equation} 
and the incremental incompressibility condition  Eq.~\eqref{eq14-1} as
\begin{equation}
 \text{tr}\; \dot {\bm {F}}_I =\frac{\partial u_r}{\partial r}+\frac{1}{r}\left(\frac{\partial u_\theta}{\partial \theta}+u_r\right)+\frac{\partial u_z}{\partial z}=0.
\label{incremental-incompressibility}
\end{equation}
Introducing the incremental electric potential $\dot\phi$, we write the components of the incremental electric field as
\begin{equation}
\dot{E}_{lIr}=-\frac{\partial \dot \phi}{\partial r},
\qquad
\dot{E}_{lI\theta}=-\frac{1}{r}\frac{\partial \dot \phi}{\partial \theta},
\qquad
\dot{E}_{lIz}=-\frac{\partial \dot \phi}{\partial z}.
\end{equation}
According to Eqs. \eqref{eq16-1} and \eqref{eq52}, we provide the explicit expression of the incremental nominal stress $\bm{\dot{S}}_I$ and incremental electric fields $\bm{\dot{E}}_{lI}$ in terms of instantaneous electro-elastic moduli and incremental fields in Appendix A.

For the growing tube, the incremental equilibrium equations and incremental  Maxwell's equations reduce to 
\begin{align}\label{incremental-constitutive3}
&\frac{\partial \dot S_{Irr}}{\partial r}+\frac{1}{r}\frac{\partial \dot S_{I\theta r}}{\partial \theta}+\frac{\dot S_{Irr}-\dot S_{I\theta\theta}}{r}+\frac{\partial \dot S_{Izr}}{\partial z}=0,\notag\\
&\frac{\partial \dot S_{Ir\theta}}{\partial r}+\frac{1}{r}\frac{\partial \dot S_{I\theta \theta}}{\partial \theta}+\frac{\dot S_{I\theta r}+\dot S_{Ir\theta}}{r}+\frac{\partial \dot S_{Iz\theta}}{\partial z}=0,\notag\\
&\frac{\partial \dot S_{Irz}}{\partial r}+\frac{1}{r}\frac{\partial \dot S_{I\theta z}}{\partial \theta}+\frac{\partial \dot S_{Izz}}{\partial z}+\frac{\dot S_{Irz}}{r}=0,
\end{align}
and
\begin{equation}\label{incremental-Maxwell}
\frac{\partial \dot{\tilde {D}}_{Ir}}{\partial r}+\frac{1}{r}\left(\frac{\partial \dot{\tilde {D}}_{I\theta}}{\partial \theta}+\dot{\tilde {D}}_{Ir}\right)+\frac{\partial \dot{\tilde {D}}_{Iz}}{\partial z}=0,
\end{equation}
respectively. 
In addition, as the incremental displacements are axisymmetric and the lateral surfaces are free of tractions, the incremental boundary conditions read
\begin{align}\label{eq57}
& u_z=\dot S_{Izr}=\dot S_{Iz\theta}=0, && \text{at } z=0,l,\notag \\
& \dot S_{Irr}=\dot S_{Ir\theta}=\dot S_{Irz}=\dot \phi=0 && \text{at } r=r_i, r_o.
\end{align}


\subsection{Stroh formulation and resolution.}


We look for solutions with sinusoidal circumferential and axial variations, as 
\begin{align}\label{incremental-solutions}
&u_r=U_r(r)\text{cos}\left(m\theta\right)\text{cos}\left(kz\right), & \quad & u_\theta=U_\theta(r)\text{sin}\left(m\theta\right)\text{cos}\left(kz\right), \notag \\
&u_z=U_z(r)\text{cos}\left(m\theta\right)\text{sin}\left(kz\right), & \quad & \dot{\phi}=\Phi(r)\text{cos}\left(m\theta\right)\text{cos}\left(kz\right), \notag \\
&\dot S_{Irr}=\Sigma_{rr}(r)\text{cos}\left(m\theta\right)\text{cos}\left(kz\right), & \quad & \dot S_{Ir\theta}=\Sigma_{r\theta}(r)\text{sin}\left(m\theta\right)\text{cos}\left(kz\right), \notag \\
&\dot S_{Irz}=\Sigma_{rz}(r)\text{cos}\left(m\theta\right)\text{sin}\left(kz\right), & \quad & \dot{\tilde {D}}_{Ir}=\Delta_r(r)\text{cos}\left(m\theta\right)\text{cos}\left(kz\right),
\end{align}
where $m$, an integer, is the circumferential wavenumber, and $k={n\pi}/({\lambda_z L})$ where $n$, another integer, is the axial half wavenumber.
Then the governing equations in Eqs.~\eqref{incremental-incompressibility}, \eqref{incremental-constitutive3} and \eqref{incremental-Maxwell} can be arranged into a first-order differential system
\begin{equation}\label{Stroh}
\frac{\text{d}}{\text{d}r}\bm \eta(r)=\frac{1}{r}\bm G(r)\bm\eta(r), 
\end{equation}
where $\bm\eta(r)=\left[ \begin{matrix} \bm {\hat U} & \bm {\hat S}\end{matrix} \right]^{T}$ is the  electro-mechanical Stroh vector, with $\bm {\hat U}=\left[ \begin{matrix} U_r & U_\theta & U_z & r\Delta_r\end{matrix} \right]^{T}$ and $\bm{ \hat S}=\left[ \begin{matrix} r\Sigma_{rr} &r\Sigma_{r\theta} & r\Sigma_{rz} & \Phi 
\end{matrix} \right]^{T}$, and $\bm G$ is the so-called Stroh matrix.
It can be decomposed into the following block structure
\begin{equation}
\bm G=\left[ \begin{matrix}
 \bm G_1 & \bm G_2 \\
 \bm G_3 & \bm G_4\end{matrix} \right],
\end{equation}
where the components of the four $4\times 4$ sub-blocks $\bm G_1$, $\bm G_2$, $\bm G_3$ and $\bm G_4$ are listed in Appendix B. 

We now use  the \textit{surface impedance matrix} method to solve numerically the Stroh differential system and obtain the dispersion equation, see \cite{destrade2009bending, ciarletta2016morphology, du2018modified,su2019finite}, for details.

The $4\times 4$ conditional impedance matrix $\bm z^i(r,r_i)$ is Hermitian, and is found by integrating numerically the following  Riccati equation
\begin{equation}\label{Riccati}
\frac{\text d\bm z^i}{\text dr}=\frac{1}{r}\left(-\bm z^i \bm G_1-\bm z^i \bm G_2 \bm z^i+\bm G_3+\bm G_4 \bm z^i\right),
\end{equation}
starting from the initial condition $\bm z^i(r_i,r_i)=\bm 0$, and ending at $r = r_o$, where the target condition is that 
\begin{equation}\label{dispersion-eq}
\text{det}\; \bm z^i(r_o,r_i)=0.
\end{equation}
Then, once $\bm z^i(r,r_i)$ has been calculated, we find the displacement in the tube by integrating 
\begin{equation}\label{two_Riccati}
 \bm z^i(r_o,r_i) \bm{\hat{U}}(r_o) = \bm 0, \qquad 
 \frac{\text d}{\text dr} \bm{ \hat U}= \frac{1}{r}\bm G_1 \bm{ \hat U}+\frac{1}{r}\bm G_2 \bm z^i \bm{ \hat U}.
\end{equation}

\color{black}

\section{Numerical results for a  neo-Hookean dielectric solid}



\subsection{Large deformation of a growing tube with electro-mechanical control}


We use the ideal neo-Hookean dielectric solid \citep{zhao2007method, dorfmann2010nonlinear, dorfmann2019instabilities, su2018wrinkles, su2019finite} to model electro-mechanical effects in bio-tissues or hydrogels; its free energy density is of the form 
 \begin{equation}\label{neo-Hookean}
 \Omega=\frac{\mu}{2}\left( I_1 - 3 \right)+\frac{1}{2\varepsilon} I_5,
 \end{equation}
and its dimensionless form is
 \begin{equation}\label{neo-Hookean}
 \bar \Omega=\frac{1}{2}\left( g^{-2}\lambda^2 +g^4 \lambda^{-2}\lambda_z^{-2} +g^{-2}\lambda_z^{2} - 3 \right)+\frac{1}{2} \lambda^{-2}\lambda_z^{-2} \bar D_R^2, 
 \end{equation}
In the absence of the internal pressure, the corresponding dimensionless radial stress in Eq.~\eqref{dimensionless-stress} is 
\begin{equation}
\bar \sigma_{rr}=\frac{g}{\lambda_z}\left(\ln\frac{\lambda_i}{\lambda}+\frac{r_o^2\left(r_i^2-r^2\right)}{r^2\left(r_i^2-r_o^2\right)}\ln\frac{\lambda_o}{\lambda_i}\right),
\end{equation}
and the dimensionless voltage in Eq.~\eqref{dimensionless-voltage} is 
\begin{equation}
\bar V=-\frac{\bar r_o \lambda_i \lambda_z^{-1}\ln \bar r_o}{\bar R_o-1}
\sqrt{\frac{1}{1-\bar r_o^2} \left(g^4 \lambda_o^2-g^4\lambda_i^2+ 2 g \lambda_z \ln\frac{\lambda_i}{\lambda_o}\right)}.
\end{equation}
%

\begin{figure}[htbp]
\begin{center}
\includegraphics[width=1\textwidth]{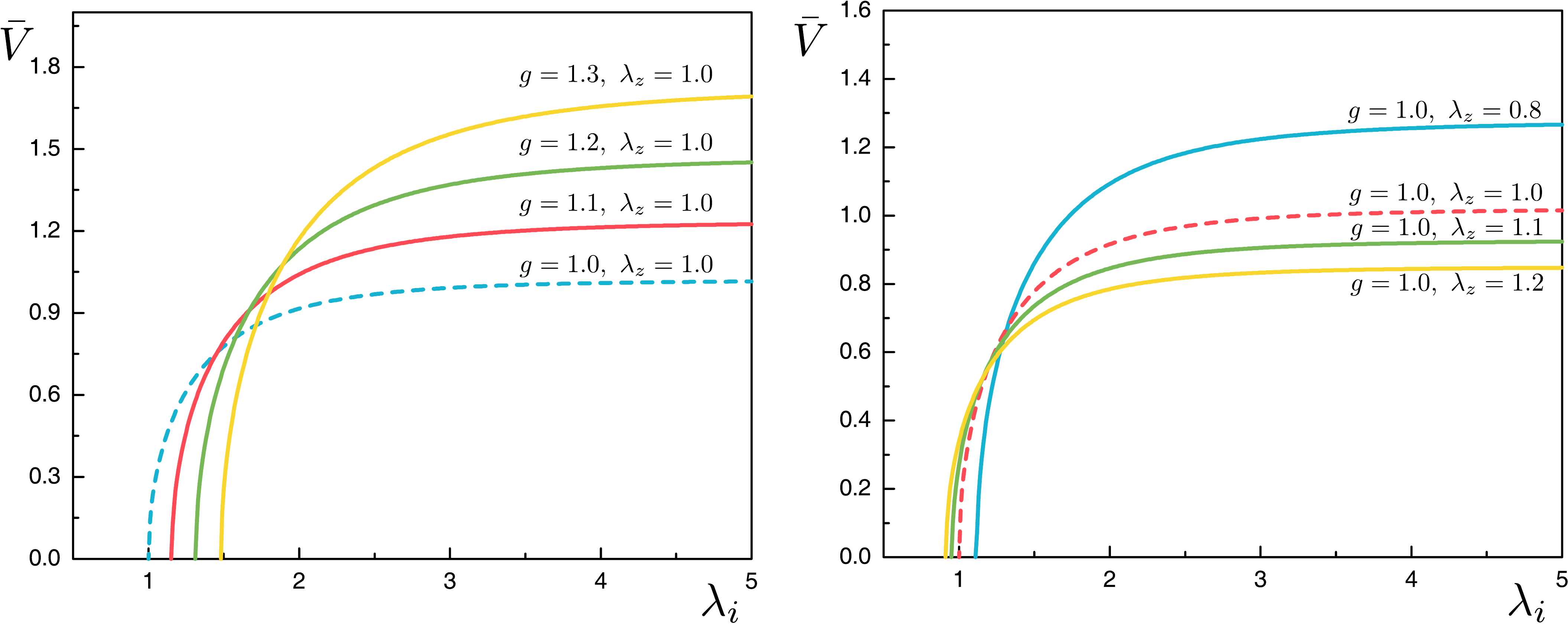}
\caption{The effect of the growth factor $g$ and axial stretch $\lambda_z$ on the nonlinear response of inner circumferential stretch $\lambda_i$ versus the applied voltage $\bar V$, in the case $\bar R_o=2.0$.}
\label{Vbar}
\end{center}
\end{figure}
Figure \ref{Vbar} shows the effects of prescribed growth and axial stretch on the nonlinear response of the inner circumferential strain $\lambda_i$ versus the applied voltage $\bar V$, see full curve and compares to the dotted line, corresponding to the no-growth scenario.
In the no-voltage case ($\bar V=0$), pure growth and contractile axial stretch lead to an increased circumferential stretch, while a tensile axial stretch leads to a reduced circumferential stretch.  
As the voltage increases ($\bar V>0$), the circumferential stretch $\lambda_i$ increases moderately until it shoots to infinity sharply for a certain threshold value of the external voltage, which is due to the absence of an axisymmetric solution of deformation \citep{shmuel2015manipulating, wu2017guided}.
In addition, we see that growth and compressive axial strain increase the threshold, while an axial stretch decreases the threshold.
%

\begin{figure}[htbp]
\begin{center}
\includegraphics[width=0.85\textwidth]{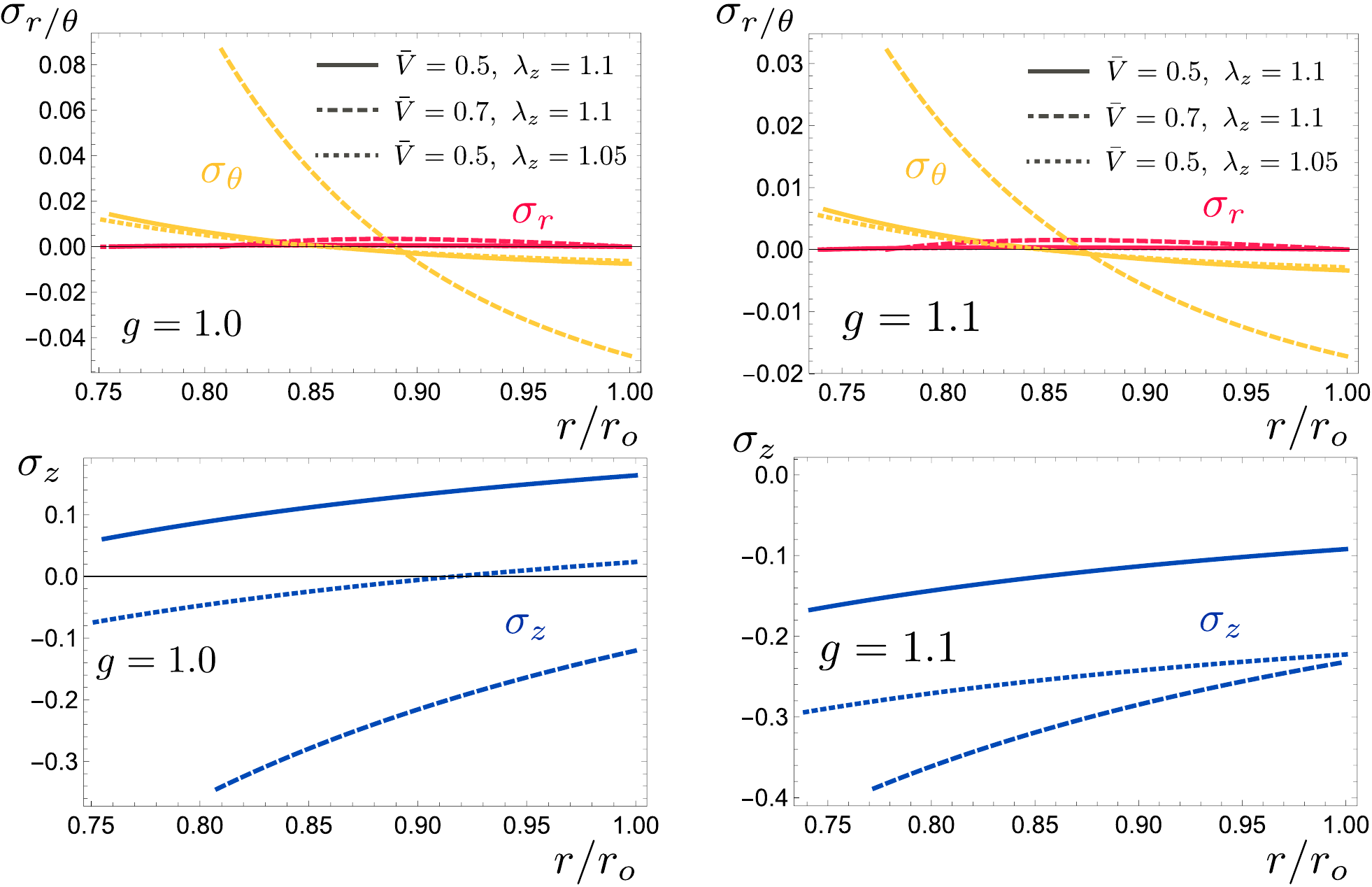}
\vspace{10pt}
\caption{The distribution of residual stress of a tube that subjects to the external electric field $\bar V$ and axial stretch $\lambda_z$ after growth, where $\bar R_o=2.0$.}
\label{RS_rthetaz}
\end{center}
\end{figure}

In Figure \ref{RS_rthetaz} we present the distribution of residual stress $\bm \sigma$ in the wall of a tube subjected to different  external electric field $\bar V$ and prescribed axial stretch $\lambda_z$, without growth ($g=1.0$), and with growth ($g=1.1$). 
Compared to the circumferential stress $\sigma_{\theta\theta}$ and the axial stress $\sigma_{zz}$, the radial stress $\sigma_{rr}$ is almost negligible in both cases.
Also, the circumferential stress $\sigma_{\theta\theta}$ is positive (compressive) at the inner face and negative (tensile) at the outer face, as expected. 
Moreover, the results show that a higher biasing voltage leads to a higher circumferential stress and a more inhomogeneous axial stress.

In the literature so far, residual stress is explained by differential growth coming either from non-homogeneous growth of multilayer structures or from anisotropic growth factors of isotropic materials.
Here it is worth noting that by applying an electrical biasing field,  residual stress can also be induced even  when growth is isotropic or homogeneous.
In addition, because  the tube  is constrained axially, the axial stress is very sensitive to both the applied axial strain and the prescribed growth factor. 
We see from the figure that with growth and voltage, the circumferential stress decreases and the axial stress can change from tensile to compressive or can be made to be more compressive. 
All these effects point to the possibility of growth and voltage thresholds of instability and pattern generation.


\subsection{Patterns formation by electro-mechanically guided growth}


The Appendix gives the dimensionless non-zero components of the instantaneous electro-elastic moduli in Eq.~\eqref{eq20-1} as
\begin{align}
&\bar{ \mathcal{A}}_{I1111}=\bar{\mathcal{A}}_{I1212}=\bar{\mathcal{A}}_{I1313}= g^4\lambda^{-2}\lambda_z^{-2}+ D_r^2,\notag\\
&\bar{\mathcal{A}}_{I2121}=\bar{\mathcal{A}}_{I2222}=\bar{\mathcal{A}}_{I2323}= g^{-2}\lambda^{2}\notag\\
&\bar{\mathcal{A}}_{I3131}=\bar {\mathcal{A}}_{I3232}=\bar{\mathcal{A}}_{I3333}= g^{-2}\lambda_z^{2}\notag\\
&\bar{\Gamma}_{I111}= 2\bar{\Gamma}_{I122}=2\bar{\Gamma}_{I133}=2\bar D_r,\notag\\
&\bar{\mathcal{K}}_{I11}=\bar{\mathcal{K}}_{I22}=\bar{\mathcal{K}}_{I33}=1.
\end{align}

Further, we may non-dimensionalise the Stroh matrix $\bm G$ as follows.
First rewrite $kr$ as
\begin{equation}\label{dimensionless_k}
kr=\bar k \dfrac{\lambda}{\bar R_o-1}\sqrt{\dfrac{\lambda_z\lambda_i^2-1}{\lambda^2\lambda_z-1}},
\end{equation}
where $\bar k = n \pi\dfrac{ R_o-R_i}{\lambda_z  L}$ and the Riccati equation in Eq.~\eqref{Riccati} as
\begin{equation}\label{Riccati_1}
\dfrac{\text d\bm z^i}{\text d\lambda}=\frac{1}{\lambda\left(1-\lambda_z\lambda^2\right)}\left(-\bm z^i \bm G_1-\bm z^i \bm G_2 \bm z^i+\bm G_3+\bm G_4 \bm z^i\right).
\end{equation}
Finally, the dispersion equation in Eq. \eqref{dispersion-eq} is  equivalent to 
\begin{equation}
\text{det}\ \bm z^i(\lambda_o,\lambda_i)=0.
\label{BFE}
\end{equation}

Then, the solution of the non-dimensional Eqs.~\eqref{Riccati_1}-\eqref{BFE} gives the critical state for an instability of the growing tube under electro-mechanical control. 
With the resulting $\bm z^i(\lambda_o,\lambda_i)$ we find the components of $\bm {\hat {U}}(r_o)$ on the outer surface by solving  Eq.~\eqref{two_Riccati}$_2$, and by integrating \eqref{two_Riccati}$_1$ we obtain the incremental displacements throughout the thickness of the tube wall, see \cite{destrade2009bending} for details.

\color{black}
In particular, our goal is to find the critical growth factor that  satisfies the target condition \eqref{dispersion-eq}. 
For a set of possible combinations of wrinkle numbers $m$ and $n$, we integrate the Riccati equation \eqref{dimensionless_k}  using a numerical differential solver (‘NDSolve’) in \textit{Mathematica}.
%
First, we iterate the growth factor $g$ until we obtain the critical growth factor $g_\text{cr}$ where the integrated solution of the Riccati equation satisfies the target condition \eqref{dispersion-eq}. 
Then to find $g_\text{cr}$  precisely, we use the bisection method.
We set the threshold of numerical accuracy to find the zero in the target condition \eqref{dispersion-eq} as being $ \leq10^{-15}$ and the step in the growth factor  as  $\delta g \leq 10^{-12}$. 
Finally, among all possible combinations of $m$ and $n$, we keep the smallest critical growth factor $g_{cr}$ as  the mode that will occur first.

\color{black}

\subsection{Pattern creation without growth}


Prior to studying electro-mechanically guided growth, we first establish the allowed ranges for external voltage and prescribed axial stretch, where the tube remains stable in the absence of growth ($g=1.0$).

\begin{figure}[ht!]
\begin{center}
\includegraphics[width=0.8\textwidth]{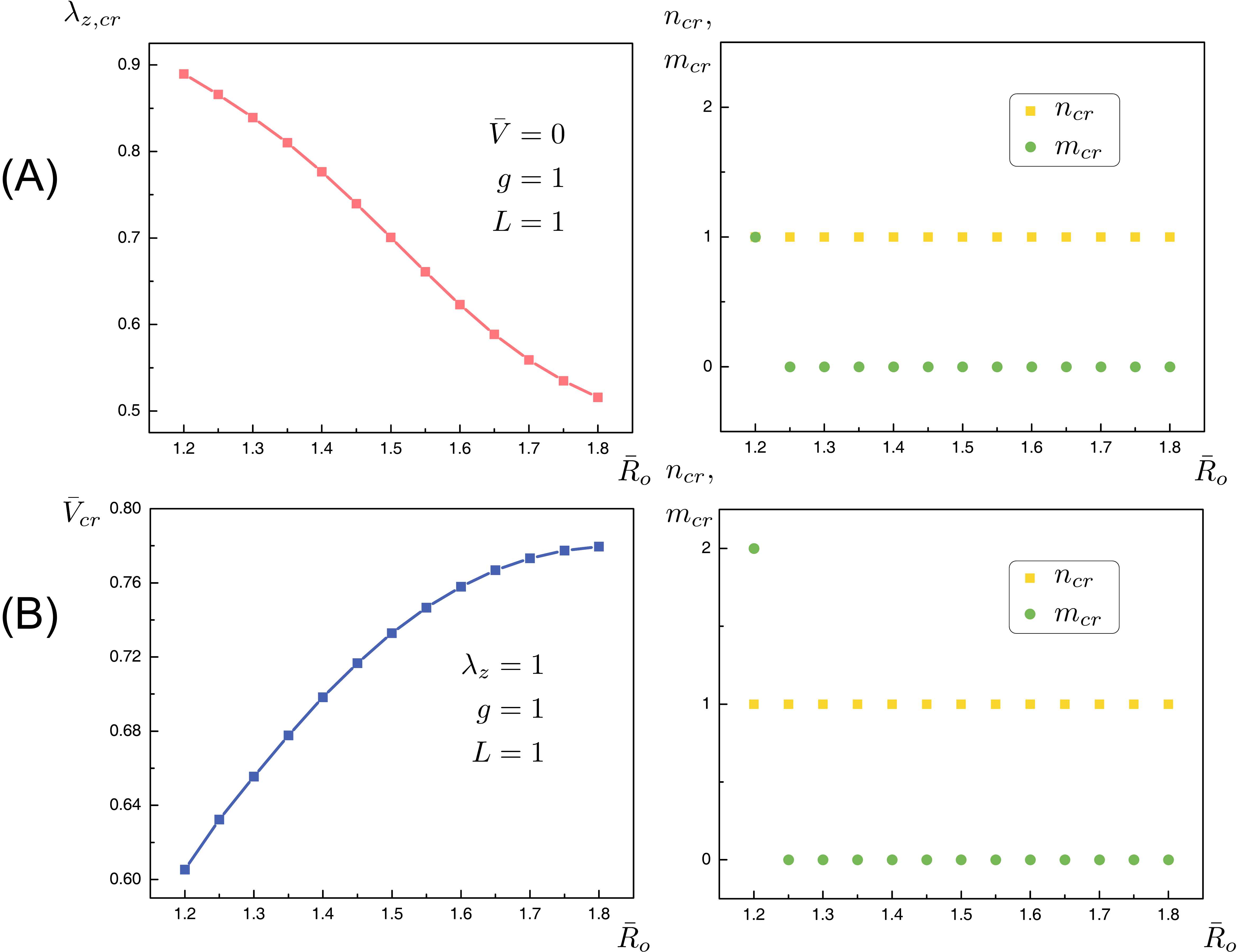}
\caption{
Tube of length $L=1.0$.
(A) Critical axial stretch $\lambda_{z, \text{cr}}$ and modes of instability $m_\text{cr}$, $n_\text{cr}$, when there is no growth ($g=1.0$) and no voltage applied ($\bar V=0.0$). 
(B). Critical voltage $\bar V_\text{cr}$ and modes of instability when there is no growth ($g=1.0)$ and no axial stretch ($\lambda_z=1.0$).}
\label{WithoutG_lambda_V}
\end{center}
\end{figure}

Figure \ref{WithoutG_lambda_V}A shows  the critical axial strain and Figure \ref{WithoutG_lambda_V}B the critical voltage for the onset of stability versus the dimensionless outer radius $\bar R_o$ when $L=1.0$. 
As expected intuitively, thicker tubes are more stable than the inner tubes, as they require larger contractile axial stretches and higher applied voltages to buckle. 
This observation is aligned with the experiment of switching crease patterns on hydrogel surfaces through low voltage performed by \citet{xu2013low}, where the critical voltage required for generating patterns is  higher for thicker blocks than for thinner blocks.
In addition, because the axial stretch is fixed, the final  patterns are almost always 2D axial buckling ($m_\text{cr}=0$, $n_\text{cr}=1$), except when the tube is thin, where  there might be simple 3D patterns emerging, with mixed axial and circumferential wrinkles (hence we can have $m_\text{cr}=n_\text{cr}=1$ when no voltage is applied, or $m_\text{cr}=2$, $n_\text{cr}=1$ when there is no axial stretch.)

\begin{figure}[ht!]
\begin{center}
\includegraphics[width=0.7\textwidth]{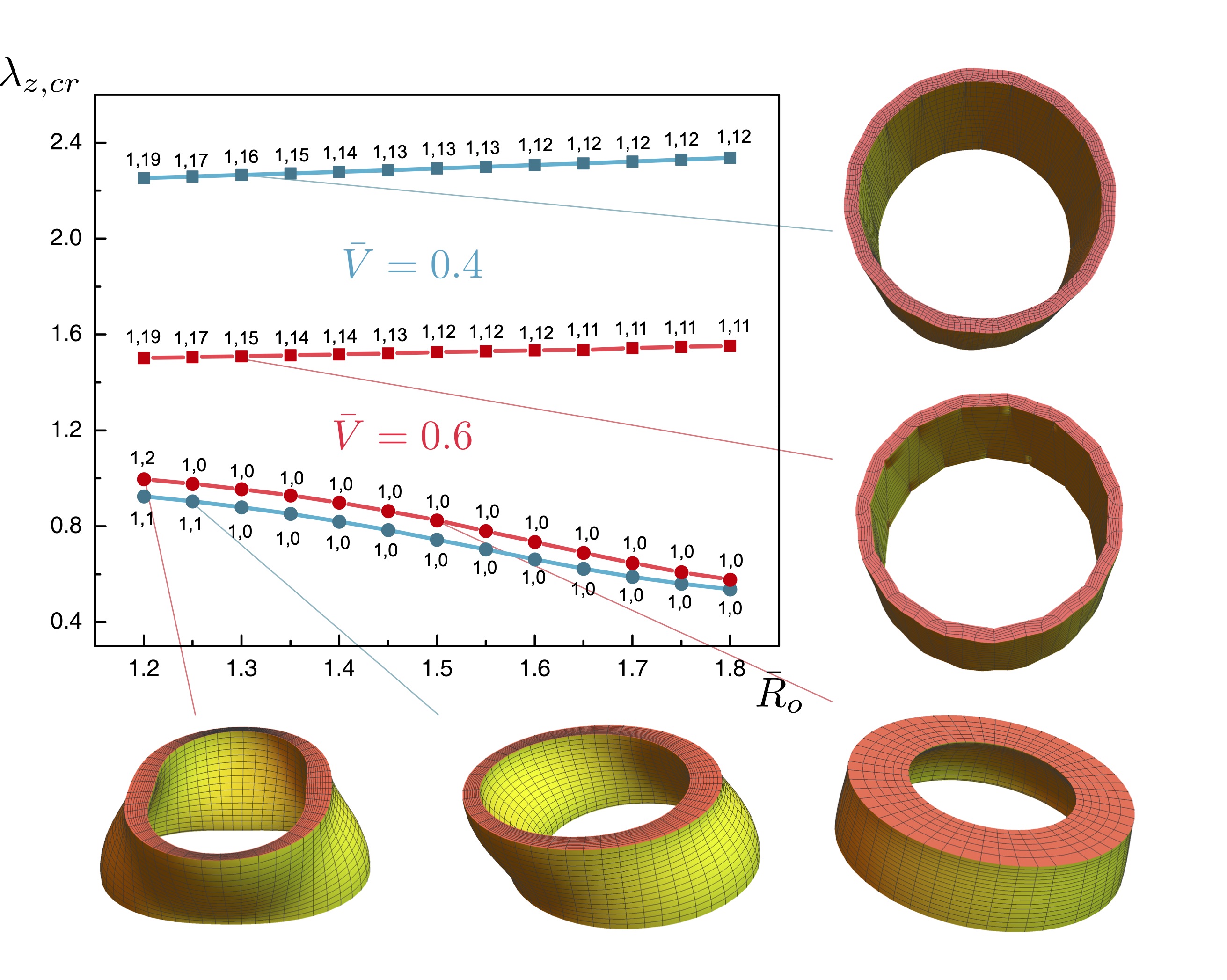}
\vspace{5pt}
\caption{Distribution of residual stress after growth subject to an external electric field $\bar V = 0.4, 0.6$ and axial stretch $\lambda_z$. 
The integer values given in the neighbourhood of each point are the couple $n_\text{cr}$, $m_\text{cr}$.}
\label{combined_bis}
\end{center}
\end{figure}

In Figure \ref{combined_bis}, we present the critical axial stretch and corresponding pattern modes obtained from a combined electro-mechanical actuation ($\lambda_z \ne 1.0$, $\bar V \ne 0.0$).
For certain applied voltages ($\bar V=0.4, 0.6$), we see that thicker tubes are again more stable than thinner tubes, as they require larger critical contractile ($\lambda_z<1$) and extensional  ($\lambda_z>1$) axial stretches to buckle. 
In contrast to the case of sole axial stretch control (when $\bar V = 0.0$), buckling may now occur \textit{in extension}, and not only in contraction.
Also, although the axial buckling mode is always $n_{cr}=1$, the circumferential number of wrinkles varies from $m_{cr}=0$ to $m_{cr}=12$ in our computations, showing many opportunities for 3D patterns.

\color{black}
\color{black}

\begin{figure}[ht!]
\begin{center}
\includegraphics[width=0.65\textwidth]{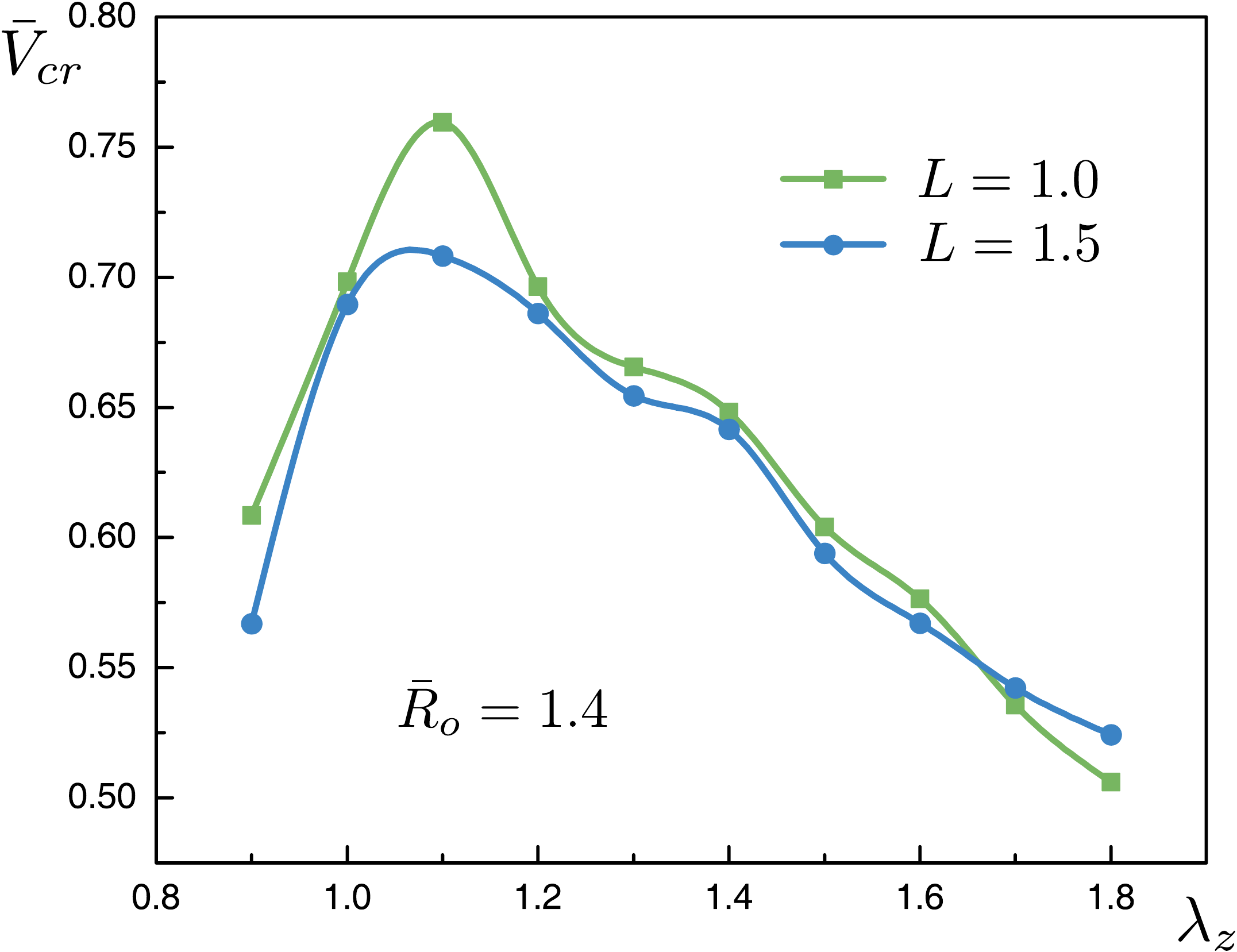}
\vspace{10pt}
\caption{The effect of axial stretch on the critical voltage with $\bar R_o=1.4$ and $L=1.0,  1.5$.}
\label{combinedlaw_bis}
\end{center}
\end{figure}

Figure \ref{combinedlaw_bis} shows the effect of the prescribed axial stretch $\lambda_z$ on the critical voltage $\bar V_\text{cr}$, for tubes with thickness measure $\bar R_o=1.4$ and heights $L=1.0, 1.5$. 
It shows a maximal critical voltage, and hence that a certain extent of axial extension can stabilise the tube by increasing the voltage controllable range; away from that value, excessive contraction or extension in the axial direction makes the tube more unstable, with a smaller controllable range.

\subsection{Pattern creation with growth}


Now, within  the controllable ranges obtained above, we investigate  instability and pattern generation caused by growth ($g \ne 1.0$) and guided by external electro-mechanical loads ($\bar V \ne 0.0$, $\lambda_z \ne 1.0$).

\begin{figure}[h!]
\begin{center}
\includegraphics[width=.9\textwidth]{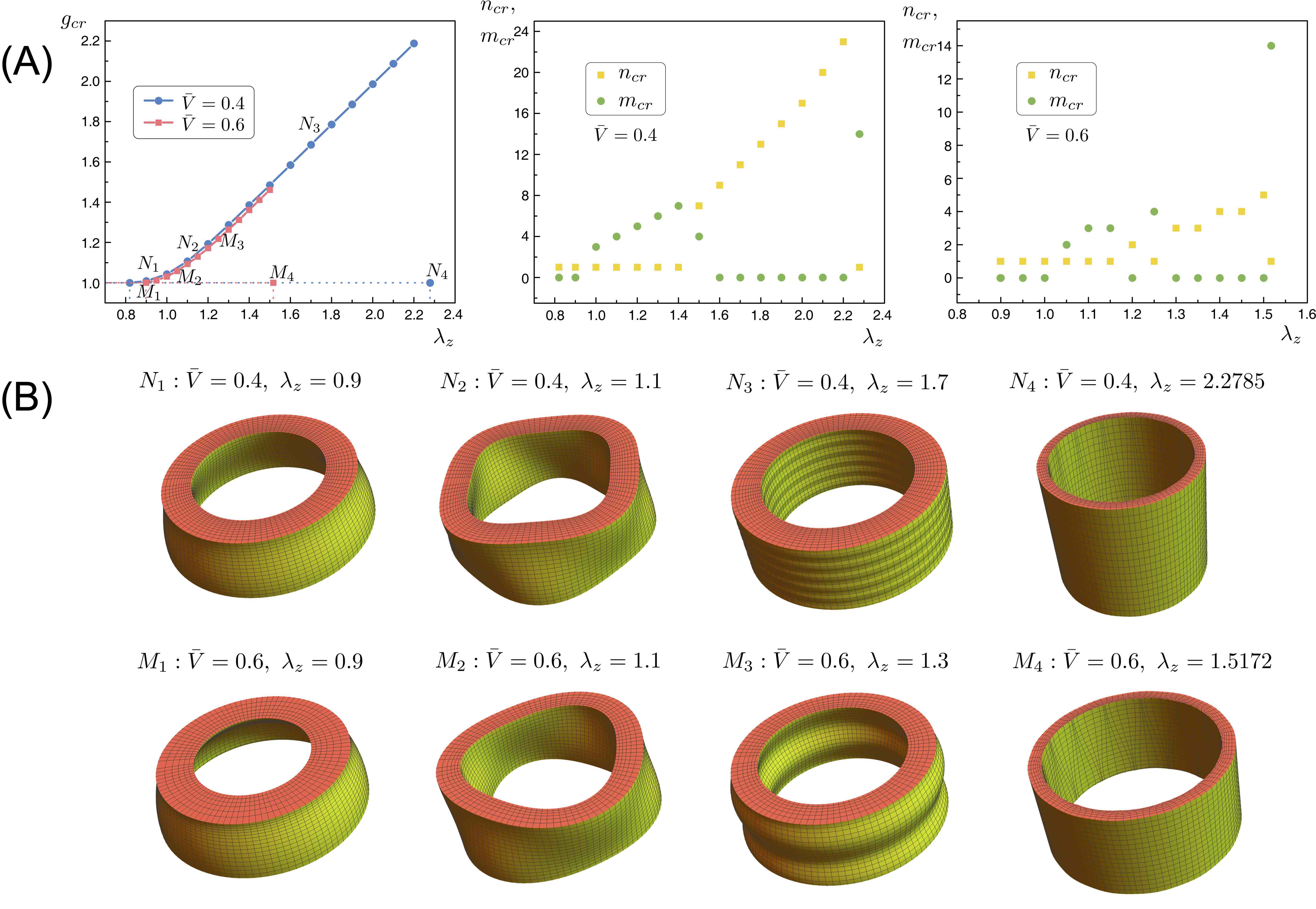}
\caption{\color{black}(A) The influence of the external axial strain on the critical growth factor and the corresponding wave numbers with applied voltage $\bar V=0.4,~0.6$, where $\bar R_0=1.4, ~R_i=1.0,~L=1.0$; 
(B) some typical patterns of (A).}
\label{ChangeLambda}
\end{center}
\end{figure}

Figure \ref{ChangeLambda} displays the effect of the prescribed axial stretch $\lambda_z$ on the critical growth factor, when the  applied voltage is $\bar V=0.4, 0.6$. 
For illustration we picked some representative points:  $N_1, \ldots, N_4$ when $\bar V = 0.4$ and $M_1, \ldots, M_4$ when $\bar V = 0.6$.

As the axial strain $\lambda_z$ increases from the critical axial compressive strain points ($N_1$ and $M_1$), the critical growth factor increases monotonically until a maximum as $\lambda_z$ reaches the critical axial tensile strain points ($N_4$ and $M_4$) obtained in the absence of growth ($g=1.0$).

In terms of shapes, the critical patterns start from a 2D axial buckling shape, and move on to various 3D mixed axial and circumferential wrinkles, to 2D axial wrinkles with high wave-numbers, and finally to the 3D mixed wrinkling shapes obtained in Figure \ref{combined_bis}.
In addition, we see that a higher external voltage ($\bar V = 0.6$) promotes an earlier onset of patterns ($0.9<\lambda_\text{z,cr} < 1.52$), but with less variety of possible shapes ($m_\text{cr} = 0,2,3,4$, $n_\text{cr} = 1,2,3,4,5$), than a lower voltage ($\bar V = 0.4$) with later onset ($0.9<\lambda_\text{z,cr} < 2.28$) but more shapes  ($m_\text{cr} = 0,3,4,5,6,7$, $n_\text{cr} = 1, 7, 9, 11, 13, 15, 17, 20, 23$).

\begin{figure}[h!]
\begin{center}
\includegraphics[width=1\textwidth]{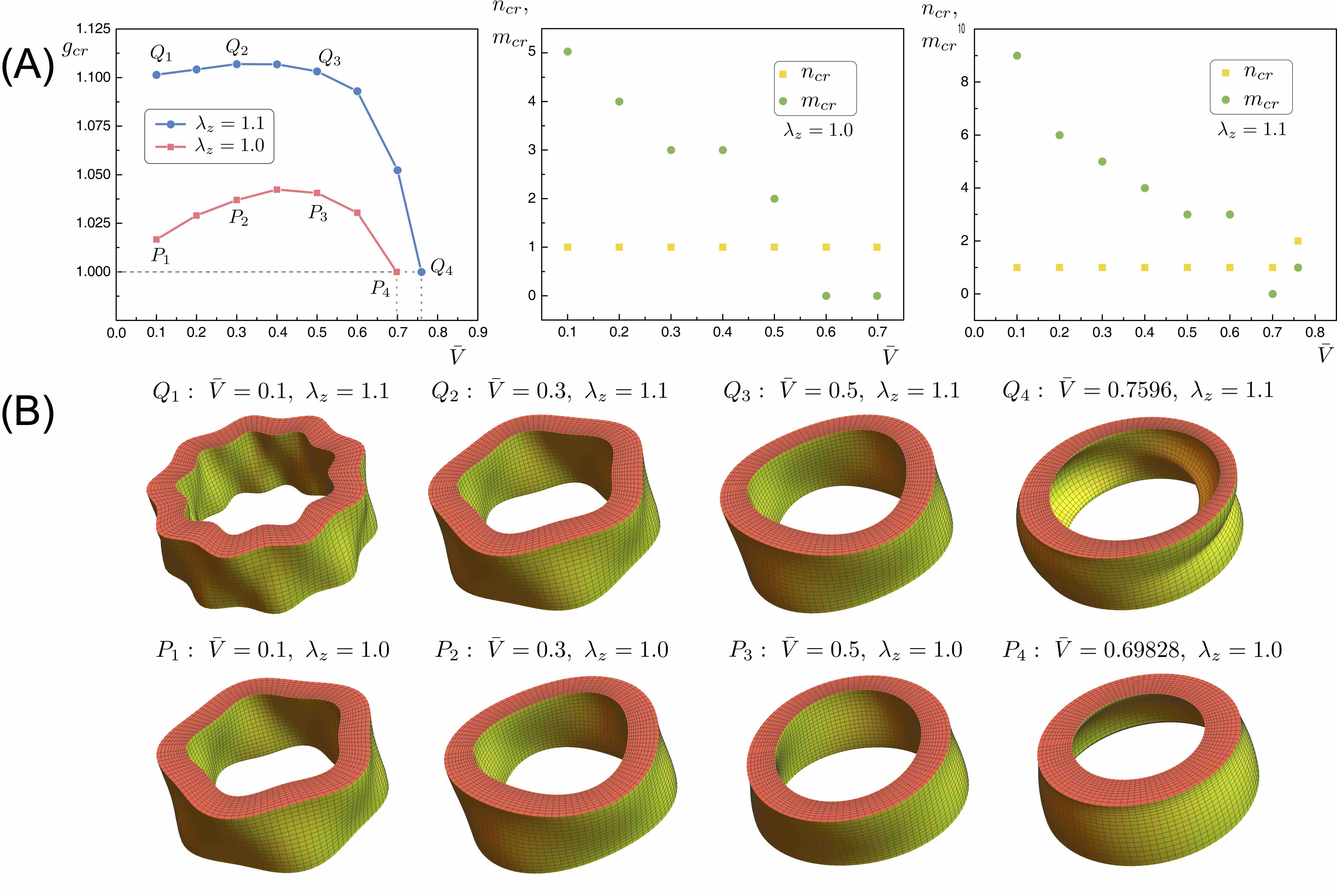}
\caption{\color{black}(A) The influence of the applied voltage on the critical growth factor and the corresponding wave numbers with axial stretches $\lambda_z=1.0,~1.1$, where $\bar R_0=1.4, ~R_i=1.0,~L=1.0$; 
(B) some typical patterns of (A).}
\label{changeV_bis}
\end{center}
\end{figure}

Finally, by imposing the axial stretch as $\lambda_z=1.0, 1.1$ in turn, we investigate the effect of the applied voltage on the critical growth factor, see  Figure \ref{changeV_bis}. 
As the external voltage increases, the critical growth factor $g_{cr}$ increases to a maximum, then decreases, and finally reaches $1.0$ at the critical voltage points obtained in Figure \ref{combinedlaw_bis} ($P_4$ and $Q_4$). The critical pattern shapes start from 3D mixed axial and circumferential wrinkles, and move on to 2D axial wrinkles or simpler 3D mixed wrinkles. 
\color{black} In addition, Figure \ref{changeV_bis} shows that as long as the critical axial tension is not reached, the buckling mode can only occur in contraction and a higher axial stretch can stabilise the tube.
 \color{black}
 

\section{Discussions and conclusions}


In this study, we established a sound framework to analyse the growth of electro-elastic materials, which are omnipresent in Nature. 
This framework allowed us to propose  a way to control or guide growth.

By assuming a multiplicative decomposition of the total growth deformation, a method which has a firm basis in volume growth theory, we went on to include growth factor and electrical displacement into the elastic deformation.
We presented a general theoretical analysis for isotropic growth and pattern formation of electro-elastic solids under external mechanical and electrical biasing fields. 
Then we conducted a linearised incremental analysis to investigate the effects of external electro-mechanical fields on the generation of growth-induced patterns. 
As an illustrative example, we provided  a three-dimensional deformation and stability analysis for the isotropic growth of a tubular structure under a fixed axial stretch and a prescribed  external voltage in the radial direction.

We first studied growth and pattern formation when external voltage and mechanical loads are present.
For isotropic growth, we revealed that the biasing electric field plays a significant role in causing the non-uniformity of the residual stress and in promoting extensional buckling.
In addition, we found that there is a maximum of the applied voltage for a certain axial stretch and growth factor, which corresponds to the symmetric collapse of the structure. 

Moreover, in the absence of growth, we found a critical value of the applied voltage that would cause wrinkling pattern formation.
The critical value is also the threshold of the controllable range of the applied voltage for guiding the growth process.
Similarly, there are two thresholds of axial strain for a certain applied voltage, which correspond to  contractile and extensional buckling, and determine the upper and lower limits, respectively, of the controllable axial stretch range for guiding growth.

In the presence of growth, the numerical results show that a higher voltage can enhance the non-uniformity of the residual stress distribution and induce extensional buckling, while a lower voltage can delay the appearance of morphology while producing more complex shapes.
Within a controllable range, axial tensile stretching shows the ability to stabilise the tube and help form more complex 3D patterns, while axial contractile stretch promotes instability.
Both the applied voltage and the prescribed axial stretch have a significant impact and a promising future on guiding the growth and patterns formation.


\color{black}
Our numerical results suggest that  growth instability and pattern formation can be guided or controlled by an electric field, instead of by purely mechanical means (such as changing elasticity, thickness, and initial residual stress, see \cite {ciarletta2014pattern, balbi2015morphoelastic, du2019prescribing, du2019influence}). \color{black} 
In principle, the coupling can be also used to design a pattern prescription strategy, growth self-assembly, drug delivery devices, or 4D bio-mimetic printing in engineering. 
However, this work only considers the dielectric characteristics that could reflect the effect of external voltage on the growth process, which may not be enough to recapture the actual electro-mechanical growth process. 
As these growable soft matters actually endow more complex electro-mechanical properties, it is, therefore, worthwhile to conduct further research in consideration of bio-piezoelectric and pyroelectric effects.

\color{black} 
On the other hand, as our results suggest that  external forces or electric fields can affect the formation of patterns, it follows that these electro-mechanical factors may also create some nonlinear interplay in the evolution of post-buckling patterns, such as  formation of creases, symmetry breaking, wrinkle mode transition, period-doubling, etc. 
Covering these phenomena requires nonlinear incremental analysis or nonlinear finite element simulations.
These challenges are interesting for future works, especially in the nonlinear stability analysis of growing electro-elastic materials, similar to the work of \citet{cai1999imperfection} on the weakly nonlinear analysis of an elastic half-space, and that of \citet{jin2019post} on a semi-analytical approach to the post-buckling analysis of elastic growth.
In addition, as there is no experimental basis showing the nature composition of the electro-mechanical growth deformation, we just choose the energy function of ideal electro-elastic solid to describe qualitatively the electric effects, which might be not sufficiently exact for certain materials. 
Therefore, more experiments about the influence of electro-mechanical factors on the complete growth process, including aspects such as growth rate and growable volume, are very welcome in the further studies.\color{black}  

%


\section*{Acknowledgement}


We gratefully acknowledge  support from the National Natural Science Foundation of China (grants 11925206/11772295) and from the China Scholarship Council.


\bibliographystyle{agsm}

@article{jin2019post,
  title={Post-buckling analysis on growing tubular tissues: A semi-analytical approach and imperfection sensitivity},
  author={Jin, Lishuai and Liu, Yang and Cai, Zongxi},
  journal={International Journal of Solids and Structures},
  volume={162},
  pages={121--134},
  year={2019},
  publisher={Elsevier}
}

@article{cai1999imperfection,
  title={On the imperfection sensitivity of a coated elastic half-space},
  author={Cai, Zongxi and Fu, Yibin},
  journal={Proceedings of the Royal Society of London. Series A: Mathematical, Physical and Engineering Sciences},
  volume={455},
  number={1989},
  pages={3285--3309},
  year={1999},
  publisher={The Royal Society}
}

@article{Choi2020,
	Author = {Choi, Moon-Young and Shin, Yerin and Lee, Hu Seung and Kim, So Yeon and Na, Jun-Hee},
	Da = {2020/02/12},
	Id = {Choi2020},
	Isbn = {2045-2322},
	Journal = {Scientific Reports},
	Number = {1},
	Pages = {2482},
	Title = {Multipolar spatial electric field modulation for freeform electroactive hydrogel actuation},
	Ty = {JOUR},
	Volume = {10},
	Year = {2020}}

@article{zhao2007method,
  title={Method to analyze electromechanical stability of dielectric elastomers},
  author={Zhao, Xuanhe and Suo, Zhigang},
  journal={Applied Physics Letters},
  volume={91},
  number={6},
  pages={061921},
  year={2007},
  publisher={American Institute of Physics}
}

@article{palleau2013reversible,
  title={Reversible patterning and actuation of hydrogels by electrically assisted ionoprinting},
  author={Palleau, Etienne and Morales, Daniel and Dickey, Michael D and Velev, Orlin D},
  journal={Nature communications},
  volume={4},
  number={1},
  pages={1--7},
  year={2013},
  publisher={Nature Publishing Group}
}

@article{wang2014cephalopod,
  title={Cephalopod-inspired design of electro-mechano-chemically responsive elastomers for on-demand fluorescent patterning},
  author={Wang, Qiming and Gossweiler, Gregory R and Craig, Stephen L and Zhao, Xuanhe},
  journal={Nature communications},
  volume={5},
  number={1},
  pages={1--9},
  year={2014},
  publisher={Nature Publishing Group}
}

@article{wang2012dynamic,
  title={Dynamic electrostatic Lithography: Multiscale on-demand patterning on large-area curved surfaces},
  author={Wang, Qiming and Tahir, Mukarram and Zang, Jianfeng and Zhao, Xuanhe},
  journal={Advanced Materials},
  volume={24},
  number={15},
  pages={1947--1951},
  year={2012},
  publisher={Wiley Online Library}
}

@article{wang2011creasing,
  title={Creasing to cratering instability in polymers under ultrahigh electric fields},
  author={Wang, Qiming and Zhang, Lin and Zhao, Xuanhe},
  journal={Physical review letters},
  volume={106},
  number={11},
  pages={118301},
  year={2011},
  publisher={APS}
}

@article{wang2011electro,
  title={Electro-creasing instability in deformed polymers: experiment and theory},
  author={Wang, Qiming and Tahir, Mukarram and Zhang, Lin and Zhao, Xuanhe},
  journal={Soft Matter},
  volume={7},
  number={14},
  pages={6583--6589},
  year={2011},
  publisher={Royal Society of Chemistry}
}

@article{xu2013low,
  title={Low-voltage switching of crease patterns on hydrogel surfaces},
  author={Xu, Bin and Hayward, Ryan C},
  journal={Advanced Materials},
  volume={25},
  number={39},
  pages={5555--5559},
  year={2013},
  publisher={Wiley Online Library}
}

@article{chae2018review,
  title={Review on electromechanical coupling properties of biomaterials},
  author={Chae, Inseok and Jeong, Chang Kyu and Ounaies, Zoubeida and Kim, Seong H},
  journal={ACS Applied Bio Materials},
  volume={1},
  number={4},
  pages={936--953},
  year={2018},
  publisher={ACS Publications}
}

@article{wieland2015investigation,
  title={Investigation of the inverse piezoelectric effect of trabecular bone on a micrometer length scale using synchrotron radiation},
  author={Wieland, DCF and Krywka, C and Mick, E and Willumeit-R{\"o}mer, R and Bader, R and Kluess, D},
  journal={Acta biomaterialia},
  volume={25},
  pages={339--346},
  year={2015},
  publisher={Elsevier}
}

@article{funk2009electromagnetic,
  title={Electromagnetic effects--From cell biology to medicine},
  author={Funk, Richard HW and Monsees, Thomas and {\"O}zkucur, Nurdan},
  journal={Progress in histochemistry and cytochemistry},
  volume={43},
  number={4},
  pages={177--264},
  year={2009},
  publisher={Elsevier}
}

@inproceedings{levin2009bioelectric,
  title={Bioelectric mechanisms in regeneration: unique aspects and future perspectives},
  author={Levin, Michael},
  booktitle={Seminars in cell \& developmental biology},
  volume={20},
  number={5},
  pages={543--556},
  year={2009},
  organization={Elsevier}
}

@article{yang2011improving,
  title={Improving protein fold recognition and template-based modeling by employing probabilistic-based matching between predicted one-dimensional structural properties of query and corresponding native properties of templates},
  author={Yang, Yuedong and Faraggi, Eshel and Zhao, Huiying and Zhou, Yaoqi},
  journal={Bioinformatics},
  volume={27},
  number={15},
  pages={2076--2082},
  year={2011},
  publisher={Oxford University Press}
}

@article{bosnjak2020experiments,
  title={Experiments and modeling of the viscoelastic behavior of polymeric gels},
  author={Bosnjak, Nikola and Nadimpalli, Siva and Okumura, Dai and Chester, Shawn A},
  journal={Journal of the Mechanics and Physics of Solids},
  volume={137},
  pages={103829},
  year={2020},
  publisher={Elsevier}
}

@article{kim2002electric,
  title={Electric stimuli responses to poly (vinyl alcohol)/chitosan interpenetrating polymer network hydrogel in NaCl solutions},
  author={Kim, Seon Jeong and Park, Sang Jun and Kim, In Young and Shin, Mi-Seon and Kim, Sun I},
  journal={Journal of applied polymer science},
  volume={86},
  number={9},
  pages={2285--2289},
  year={2002},
  publisher={Wiley Online Library}
}

@article{agnihotri2005electrically,
  title={Electrically modulated transport of diclofenac salts through hydrogels of sodium alginate, carbopol, and their blend polymers},
  author={Agnihotri, Sunil A and Kulkarni, Raghavendra V and Mallikarjuna, Nadagouda N and Kulkarni, Padmakar V and Aminabhavi, Tejraj M},
  journal={Journal of applied polymer science},
  volume={96},
  number={2},
  pages={301--311},
  year={2005},
  publisher={Wiley Online Library}
}

@article{shmuel2015manipulating,
  title={Manipulating torsional motions of soft dielectric tubes},
  author={Shmuel, Gal},
  journal={Journal of Applied Physics},
  volume={117},
  number={17},
  pages={174902},
  year={2015},
  publisher={AIP Publishing LLC}
}

@article{wu2017guided,
  title={On guided circumferential waves in soft electroactive tubes under radially inhomogeneous biasing fields},
  author={Wu, Bin and Su, Yipin and Chen, Weiqiu and Zhang, Chuanzeng},
  journal={Journal of the Mechanics and Physics of Solids},
  volume={99},
  pages={116--145},
  year={2017},
  publisher={Elsevier}
}

@article{ciarletta2016morphology,
  title={Morphology of residually stressed tubular tissues: Beyond the elastic multiplicative decomposition},
  author={Ciarletta, Pasquale and Destrade, M and Gower, AL and Taffetani, Matteo},
  journal={Journal of the Mechanics and Physics of Solids},
  volume={90},
  pages={242--253},
  year={2016},
  publisher={Elsevier}
}

@article{destrade2009bending,
  title={Bending instabilities of soft biological tissues},
  author={Destrade, Michel and Annaidh, Aisling Ni and Coman, Ciprian D},
  journal={International Journal of Solids and Structures},
  volume={46},
  number={25-26},
  pages={4322--4330},
  year={2009},
  publisher={Elsevier}
}

@article{su2019finite,
  title={Finite bending and pattern evolution of the associated instability for a dielectric elastomer slab},
  author={Su, Yipin and Wu, Bin and Chen, Weiqiu and Destrade, Michel},
  journal={International Journal of Solids and Structures},
  volume={158},
  pages={191--209},
  year={2019},
  publisher={Elsevier}
}

@article{su2018wrinkles,
  title={Wrinkles in soft dielectric plates},
  author={Su, Yipin and Broderick, Hannah Conroy and Chen, Weiqiu and Destrade, Michel},
  journal={Journal of the Mechanics and Physics of Solids},
  volume={119},
  pages={298--318},
  year={2018},
  publisher={Elsevier}
}

@article{dorfmann2005nonlinear,
  title={Nonlinear electroelasticity},
  author={Dorfmann, A and Ogden, RW},
  journal={Acta Mechanica},
  volume={174},
  number={3-4},
  pages={167--183},
  year={2005},
  publisher={Springer}
}

@article{dorfmann2019instabilities,
  title={Instabilities of soft dielectrics},
  author={Dorfmann, Luis and Ogden, Ray W},
  journal={Philosophical Transactions of the Royal Society A},
  volume={377},
  number={2144},
  pages={20180077},
  year={2019},
  publisher={The Royal Society Publishing}
}

@article{dorfmann2010nonlinear,
  title={Nonlinear electroelastostatics: Incremental equations and stability},
  author={Dorfmann, Alois and Ogden, Ray W},
  journal={International Journal of Engineering Science},
  volume={48},
  number={1},
  pages={1--14},
  year={2010},
  publisher={Elsevier}
}

@article{zelisko2015mechanism,
  title={What is the mechanism behind biological ferroelectricity?},
  author={Zelisko, Matthew and Li, Jiangyu and Sharma, Pradeep},
  journal={Extreme Mechanics Letters},
  volume={4},
  pages={162--174},
  year={2015},
  publisher={Elsevier}
}

@article{liu2014ferroelectric,
  title={Ferroelectric switching of elastin},
  author={Liu, Yuanming and Cai, Hong-Ling and Zelisko, Matthew and Wang, Yunjie and Sun, Jinglan and Yan, Fei and Ma, Feiyue and Wang, Peiqi and Chen, Qian Nataly and Zheng, Hairong and others},
  journal={Proceedings of the National Academy of Sciences},
  volume={111},
  number={27},
  pages={E2780--E2786},
  year={2014},
  publisher={National Acad Sciences}
}

@article{jaffe1984electric,
  title={Electric fields and wound healing},
  author={Jaffe, Lionel F and Vanable Jr, Joseph W},
  journal={Clinics in dermatology},
  volume={2},
  number={3},
  pages={34--44},
  year={1984},
  publisher={Elsevier}
}

@article{jaffe1977electrical,
  title={Electrical controls of development},
  author={Jaffe, Lionel F and Nuccitelli, Richard},
  journal={Annual review of biophysics and bioengineering},
  volume={6},
  number={1},
  pages={445--476},
  year={1977},
  publisher={Annual Reviews 4139 El Camino Way, PO Box 10139, Palo Alto, CA 94303-0139, USA}
}

@article{Fukada1957,
author = {Fukada, Eiichi and Yasuda, Iwao},
journal = {Journal of the Physical Society of Japan},
number = {10},
pages = {1158--1162},
title = {{On the Piezoelectric Effect of Bone}},
volume = {12},
year = {1957}
}

@book{martin1998skeletal,
  title={Skeletal tissue mechanics},
  author={Martin, R Bruce and Burr, David B and Sharkey, Neil A and Fyhrie, David P},
  volume={190},
  year={1998},
  publisher={Springer}
}

@article{Fung1991,
author = {Fung, Y C},
journal = {Annals of Biomedical Engineering},
number = {3},
pages = {237},
title = {{What are the residual stresses doing in our blood vessels?}},
volume = {19},
year = {1991}
}

@article{ciarletta2014pattern,
  title={Pattern selection in growing tubular tissues},
  author={Ciarletta, Pasquale and Balbi, Valentina and Kuhl, E},
  journal={Physical review letters},
  volume={113},
  number={24},
  pages={248101},
  year={2014},
  publisher={APS}
}

@article{rodriguez1994stress,
  title={Stress-dependent finite growth in soft elastic tissues},
  author={Rodriguez, Edward K and Hoger, Anne and McCulloch, Andrew D},
  journal={Journal of biomechanics},
  volume={27},
  number={4},
  pages={455--467},
  year={1994},
  publisher={Elsevier}
}

@article{anderson1968electrical,
  title={Electrical properties of wet collagen},
  author={Anderson, JC and Eriksson, C},
  journal={Nature},
  volume={218},
  number={5137},
  pages={166},
  year={1968},
  publisher={Nature Publishing Group}
}

@article{athenstaedt1970permanent,
  title={Permanent longitudinal electric polarization and pyroelectric behaviour of collagenous structures and nervous tissue in man and other vertebrates},
  author={Athenstaedt, Herbert},
  journal={Nature},
  volume={228},
  number={5274},
  pages={830},
  year={1970},
  publisher={Nature Publishing Group}
}

@article{fukada1969piezoelectric,
  title={Piezoelectric effect in blood vessel walls},
  author={Fukada, Eiichi and Hara, Kiyoshi},
  journal={Journal of the Physical Society of Japan},
  volume={26},
  number={3},
  pages={777--780},
  year={1969},
  publisher={The Physical Society of Japan}
}

@article{marino1970piezoelectric,
  title={Piezoelectric effect and growth control in bone},
  author={Marino, Andrew A and Becker, Robert O},
  journal={Nature},
  volume={228},
  number={5270},
  pages={473--474},
  year={1970},
  publisher={Springer}
}

@article{liu2012biological,
  title={Biological ferroelectricity uncovered in aortic walls by piezoresponse force microscopy},
  author={Liu, Yuanming and Zhang, Yanhang and Chow, Ming-Jay and Chen, Qian Nataly and Li, Jiangyu},
  journal={Physical review letters},
  volume={108},
  number={7},
  pages={078103},
  year={2012},
  publisher={APS}
}

@inproceedings{zhao2009electrical,
  title={Electrical fields in wound healing - an overriding signal that directs cell migration},
  author={Zhao, Min},
  booktitle={Seminars in cell developmental biology},
  volume={20},
  number={6},
  pages={674--682},
  year={2009},
  organization={Elsevier}
}

@book{goriely2017mathematics,
  title={The mathematics and mechanics of biological growth},
  author={Goriely, Alain},
  volume={45},
  year={2017},
  publisher={Springer}
}

@article{lewis2008signals,
  title={From signals to patterns: space, time, and mathematics in developmental biology},
  author={Lewis, Julian},
  journal={Science},
  volume={322},
  number={5900},
  pages={399--403},
  year={2008},
  publisher={American Association for the Advancement of Science}
}

@article{mendoncca2003directly,
  title={Directly applied low intensity direct electric current enhances peripheral nerve regeneration in rats},
  author={Mendon{\c{c}}a, Adriana Clemente and Barbieri, Cl{\'a}udio Henrique and Mazzer, Nilton},
  journal={Journal of Neuroscience Methods},
  volume={129},
  number={2},
  pages={183--190},
  year={2003},
  publisher={Elsevier}
}

@article{ahn2009relevance,
  title={Relevance of collagen piezoelectricity to Wolff's Law: a critical review},
  author={Ahn, Andrew C and Grodzinsky, Alan J},
  journal={Medical engineering \& physics},
  volume={31},
  number={7},
  pages={733--741},
  year={2009},
  publisher={Elsevier}
}

@article{levin2014molecular,
  title={Molecular bioelectricity: how endogenous voltage potentials control cell behavior and instruct pattern regulation in vivo},
  author={Levin, Michael},
  journal={Molecular biology of the cell},
  volume={25},
  number={24},
  pages={3835--3850},
  year={2014},
  publisher={Am Soc Cell Biol}
}

@article{amar2005growth,
  title={Growth and instability in elastic tissues},
  author={Amar, Martine Ben and Goriely, Alain},
  journal={Journal of the Mechanics and Physics of Solids},
  volume={53},
  number={10},
  pages={2284--2319},
  year={2005},
  publisher={Elsevier}
}

@article{balbi2015morphoelastic,
  title={Morphoelastic control of gastro-intestinal organogenesis: theoretical predictions and numerical insights},
  author={Balbi, Valentina and Kuhl, E and Ciarletta, Pasquale},
  journal={Journal of the Mechanics and Physics of Solids},
  volume={78},
  pages={493--510},
  year={2015},
  publisher={Elsevier}
}

@article{li2011surface,
  title={Surface wrinkling of mucosa induced by volumetric growth: theory, simulation and experiment},
  author={Li, Bo and Cao, Yan-Ping and Feng, Xi-Qiao and Gao, Huajian},
  journal={Journal of the Mechanics and Physics of Solids},
  volume={59},
  number={4},
  pages={758--774},
  year={2011},
  publisher={Elsevier}
}

@article{du2019influence,
  title={Influence of initial residual stress on growth and pattern creation for a layered aorta},
  author={Du, Yangkun and L{\"u}, Chaofeng and Destrade, Michel and Chen, Weiqiu},
  journal={Scientific reports},
  volume={9},
  number={1},
  pages={8232},
  year={2019},
  publisher={Nature Publishing Group}
}

@article{du2019prescribing,
  title={Prescribing Patterns in Growing Tubular Soft Matter by Initial Residual Stress},
  author={Du, Yangkun and Lu, Chaofeng and Liu, Congshan and Han, Zilong and Li, Jian and Chen, Weiqiu and Qu, Shaoxing and Destrade, Michel},
  journal={Soft Matter},
  year={2019},
  publisher={Royal Society of Chemistry}
}

@article{du2018modified,
  title={Modified multiplicative decomposition model for tissue growth: Beyond the initial stress-free state},
  author={Du, Yangkun and L{\"u}, Chaofeng and Chen, Weiqiu and Destrade, Michel},
  journal={Journal of the Mechanics and Physics of Solids},
  volume={118},
  pages={133--151},
  year={2018},
  publisher={Elsevier}
}

\newpage
\section*{Appendix A}
\label{A}
\appendix
With the most general energy function written in terms of the five invariants, the non-zero components of the instantaneous electro-elastic moduli  read \citep{dorfmann2010nonlinear,wu2017guided,su2019finite}
\begin{align}\label{appendix1}
& \mathcal A_{01111}=2\lambda^{-4}\lambda_z^{-4}\left(g^2\lambda^2\left(\lambda^4+\lambda_z^2\right)\left(\Omega_2+4 D_r^2 \Omega_{25}\right) 
\right. 
\notag\\ 
& \phantom{\mathcal A_{xxxxx}=} 
 \left.
+4 g^6\left(\lambda^2+\lambda_z^2\right)\left(\Omega_{12} + 2 D_r^2 \Omega_{26}\right) + D_r^2 \lambda^4\lambda_z^4\left(\Omega_{5} + 2 D_r^2\Omega_{55}\right)
\right. 
\notag\\ 
&\phantom{\mathcal A_{xxxxx}=} 
 \left.
+g^4\left(2 \lambda^4 \Omega_{22}+2 \lambda_z^4 \Omega_{22}+\lambda^2\lambda_z^2\left(\Omega_{1} + 4 \Omega_{22}+8 D_r^4 \Omega_{56}+D_r^2\left(4 \Omega_{15}+6\Omega_{6}\right)\right) \right)
\right. 
\notag\\ 
& \phantom{\mathcal A_{xxxxx}=} 
\left.
+ 2 g^8 \left(\Omega_{11}+4 D_r^2 \left(\Omega_{16}+D_r^2\Omega_{66}\right)\right)\right),
\notag\\
& \mathcal A_{01122}= 4\left(\left(2+\lambda^2 \lambda_z^{-2}\right)\Omega_{12}+g^4\lambda^{-4} \lambda_z^{-2} \left(\lambda^2 + \lambda_z^2\right)\Omega_{22} +g^{-2}\left(D_r^2\lambda^2 \Omega_{15}+\left(\lambda^2+\lambda_z^2\right)\Omega_{22}\right)
\right. 
\notag\\ 
& \phantom{\mathcal A_{01111}=} 
\left.
+g^{-4}D_r^2\lambda^2\lambda_z^2\Omega_{25} + g^2 \lambda^{-2}\left(\Omega_{11} +\Omega_2 +D_r^2 \left(2\Omega_{16} +\Omega_{25} \right) \right)
\right. 
\notag\\ 
& \phantom{\mathcal A_{01111}=} 
\left.
+2D_r^2 \Omega_{26} +g^6  \lambda^{-4}\lambda_z^{-2} \left(\Omega_{12}+2 D_r^2 \Omega_{26}\right) \right),
\notag\\
& \mathcal A_{01133}= 4\left(\left(2+\lambda^{-2} \lambda_z^{2}\right)\Omega_{12}+g^4\lambda^{-4} \lambda_z^{-2} \left(\lambda^2 + \lambda_z^2\right)\Omega_{22} +g^{-2}\left(D_r^2\lambda^2 \Omega_{15}+\left(\lambda^2+\lambda_z^2\right)\Omega_{22}\right)
\right. 
\notag\\ 
& \phantom{\mathcal A_{01111}=} 
\left.
+g^{-4}D_r^2\lambda^2\lambda_z^2\Omega_{25} + g^2 \lambda^{-2}\left(\Omega_{11} +\Omega_2 +D_r^2 \left(2\Omega_{16} +\Omega_{25} \right) \right)
\right. 
\notag\\
 & \phantom{\mathcal A_{01111}=} 
 \left.
+2D_r^2 \Omega_{26} +g^6  \lambda^{-4}\lambda_z^{-2} \left(\Omega_{12}+2 D_r^2 \Omega_{26}\right)\right),
\notag\\
& \mathcal A_{01212}= 2 g^4 \lambda^{-2}\lambda_z^{-2}\left(\Omega_1+2 D_r^2 \Omega_6 +\lambda_z^2 g^{-2}\left(\Omega_2+g^{-4}D_r^2\left(g^2 \lambda^2\Omega_5+\lambda^4\Omega_6\right)\right)\right), 
\notag \\
& \mathcal A_{01313}= 2 g^4 \lambda^{-2}\lambda_z^{-2}\left(\Omega_1+2 D_r^2 \Omega_6 +\lambda g^{-2}\left(\Omega_2+g^{-4}D_r^2\left(g^2 \lambda_z^2\Omega_5+\lambda_z^4\Omega_6\right)\right)\right), 
\notag \\
& \mathcal A_{01221} = -2 g^2\lambda_z^{-2}\Omega_2+2 g^{-2} \lambda^{2} \Omega_6 D_r^2, 
\notag\\
& \mathcal A_{01331}=-2 g^2\lambda^{-2}\Omega_2+2 g^{-2} \lambda_z^{2} \Omega_6 D_r^2,
\notag \\
& \mathcal A_{02121}= 2 g^{-2} \lambda^2\left(\Omega_1+g^{-2} \lambda_z^2\Omega_2+\Omega_6 D_r^2\right), 
\notag \\
& \mathcal A_{03131}= 2 g^{-2}\lambda_z^2\left(\Omega_1+ g^{-2} \lambda^2\Omega_2+\Omega_6D_r^2\right), 
\notag \\
& \mathcal A_{02222}=2 g^{-8}\left(2g^4\lambda^4 \Omega_{11}+4 g^2 \lambda^4  \lambda_z^2 \Omega_{12}+ g^6 \lambda^2 \left(\Omega_{1}+4 \Omega_{12}\right) \right. 
\notag\\
 &\phantom{\mathcal A_{01111}=} 
  \left.
+\lambda_z^{-2} g^4 \left(g^6+\lambda^2 \lambda_z^4 \right)\Omega_{2}+\lambda_z^{-4} \left(g^6+\lambda^2 \lambda_z^4 \right)\Omega_{22},\right),
\notag \\
& \mathcal A_{02233}= 4 \left(2 \Omega_{12} + g^6 \lambda^2 \lambda_z^2 \left(\left(\lambda^2 + \lambda_z^2\right)\Omega_{12}+g^2\left(\Omega_{11}+\Omega_{2}\right)\right)
\right. 
\notag\\ 
& \phantom{\mathcal A_{01111}=} 
\left.
+g^6 \lambda^{-2} \lambda_z^{-2} \left(g^6+\lambda^4 \lambda_z^2\right)\left(g^6+\lambda^2 \lambda_z^4\right)\Omega_{22}\right),
 \notag \\
& \mathcal A_{02323}=2g^{-2}\lambda^2\Omega_1+2g^2\lambda_z^{-2}\Omega_2, 
\notag \\
&\mathcal A_{02332}=-2g^{-4}\lambda^2\lambda_z^{2}\Omega_2, 
\notag \\
& \mathcal A_{03232}= 2g^{-2}\lambda_z^2\Omega_1+2 g^{2} \lambda^{-2}\Omega_2, 
\notag \\
& \mathcal A_{03333}= 2 g^{-8}\left( 2 g^4 \lambda_z^4 \Omega_{11}+4g^2\lambda^2\lambda_z^4\Omega_{12}+g^6 \lambda_z^2 \left(\Omega_{1} + 4\Omega_{12}\right)
\right. 
\notag\\ 
& \phantom{\mathcal A_{01111}=} 
\left.
+g^4\lambda^{-2}\left(g^6+\lambda^4\lambda_z^2\right)\Omega_{2}+2\lambda^4\left(g^6+\lambda^4\lambda_z^2\right)^2\Omega_{22}\right),
\end{align}
\begin{align}
\Gamma_{0111}=& 4 g^{-4}\lambda^{-4}\lambda_z^{-4}D_r\left(g^2 \lambda^4 \lambda_z^4 \left(\lambda^2+ \lambda_z^2\right)\Omega_{24}+g^6\lambda^2 \lambda_z^2\left(\lambda^2+ \lambda_z^2\right)\Omega_{25}
\right. \notag\\ & \left.
+g^{10}\left(\lambda^2+ \lambda_z^2\right)\Omega_{26}+D_r^2\lambda^{6}\lambda_z^{6}\Omega_{45}+g^4\lambda^4\lambda_z^4\left(\Omega_{14} +\Omega_{5}+D_r^2\left(2 \Omega_{46}+\Omega_{55}\right)\right)
\right. \notag\\ & \left.
+g^8\lambda^{2}\lambda_z^{2}\left(\Omega_{15}+3D_r^2 \Omega_{56}+2\Omega_{6}\right)+g^{12}\left(\Omega_{16}+2 D_r^2\Omega_{66} \right)
\right),
 \notag\\
\Gamma_{0122}=& 2 D_r\left(\Omega_5+g^{-2}\lambda^{-2}\left(\lambda^4+g^6\lambda_z^{-2}\right)\Omega_6\right), \notag \\
\Gamma_{0133}=& 2 D_r\left(\Omega_5+g^{-2}\lambda_z^{-2}\left(\lambda_z^4+g^6\lambda^{-2}\right)\Omega_6\right),\notag \\
\Gamma_{0221}=& 4g^{-6}D_r\left(\lambda^{4}\lambda_z^{2}\left(\Omega_{14}+g^{-2}\lambda_z^{2}\Omega_{24}\right)
+g^8 \lambda_z^{-2}\left(\Omega_{16}+\Omega_{25}\right)
\right. \notag\\ & \left.
+g^2 \lambda^2\left(g^2\left(\Omega_{15}+\Omega_{24}\right)+\lambda_z^2 \Omega_{25}\right)+g^6 \Omega_{26} +g^{12}\lambda^{-2}\lambda_z^{-4}\Omega_{26}\right),
\notag\\
\Gamma_{0331}=& 4g^{-6}D_r\left(\lambda_z^{4}\lambda^{2}\left(\Omega_{14}+g^{-2}\lambda^2\Omega_{24}\right)
+g^8 \lambda^{-2}\left(\Omega_{16}+\Omega_{25}\right)
\right. \notag\\ & \left.
+g^2 \lambda_z^2\left(g^2\left(\Omega_{15}+\Omega_{24}\right)+\lambda^2 \Omega_{25}\right)+g^6 \Omega_{26} +g^{12}\lambda_z^{-2}\lambda
^{-4}\Omega_{26}\right),
\end{align}
\begin{align}
\mathcal{K}_{011}=&2\left(2g^{-8} D_r^2\lambda^4\lambda_z^4 \Omega_{44} +g^{-4} \lambda^2\lambda_z^2 \left(\Omega_{4}+4 D_r^2 \Omega_{45}\right)+\Omega_5
+2D_r^2\left(2\Omega_{46}+\Omega_{55}\right)
\right. \notag\\ & \left.
+g^4\lambda^{-2}\lambda_z^{-2}\left(4 D_r^2 \Omega_{56}+\Omega_{6}\right)+2\lambda^{-4}\lambda_z^{-4} D_r^2 g^8 \Omega_{66}
\right),
\notag\\
\mathcal K_{022}=&2\left(g^2\lambda^{-2} \Omega_4 +\Omega_5 +g^{-2}\lambda^2 \Omega_6\right),
\notag\\
\mathcal K_{033}=&2\left(g^2\lambda_z^{-2} \Omega_4 +\Omega_5 +g^{-2}\lambda_z^2 \Omega_6\right).
\end{align}
\color{black}
Further, we can write the components of the updated incremental nominal stress $\bm{\dot{S}}_I$ in \eqref{eq16-1}$_1$ as
 \begin{align}\label{incremental-constitutive1}
&\dot S_{Irr}=\left(\mathcal{A}_{I1111}+p\right)\frac{\partial u_r}{\partial r}+\mathcal{A}_{I1122}\frac{1}{r}\left(\frac{\partial u_\theta}{\partial \theta}+u_r\right)+\mathcal{A}_{I1133}\frac{\partial u_z}{\partial z}+\Gamma_{I111} \dot{\tilde {D}}_{Ir} -\dot p,\notag \\
&\dot S_{I\theta\theta}=\mathcal{A}_{I1122}\frac{\partial u_r}{\partial r}+\left(\mathcal{A}_{I2222}+p\right)\frac{1}{r}\left(\frac{\partial u_\theta}{\partial \theta}+u_r\right)+\mathcal{A}_{I2233}\frac{\partial u_z}{\partial z}+\Gamma_{I221}\dot{\tilde {D}}_{Ir}-\dot p,\notag \\
&\dot S_{Izz}=\mathcal{A}_{I1133}\frac{\partial u_r}{\partial r}+\mathcal{A}_{I2233}\frac{1}{r}\left(\frac{\partial u_\theta}{\partial \theta}+u_r\right)+\left(\mathcal{A}_{I3333}+p\right)\frac{\partial u_z}{\partial z}+\Gamma_{I331}\dot{\tilde {D}}_{Ir}-\dot p,\notag \\
&\dot S_{Ir\theta}=\mathcal{A}_{I1212}\frac{\partial u_\theta}{\partial r}+\left(\mathcal{A}_{I1221}+p\right)\frac{1}{r}\left(\frac{\partial u_r}{\partial \theta}-u_\theta\right)+\Gamma_{I122}\dot{\tilde {D}}_{I\theta},\notag \\
&\dot S_{Irz}=\mathcal{A}_{I1313}\frac{\partial u_z}{\partial r}+\left(\mathcal{A}_{I1331}+p\right)\frac{\partial u_r}{\partial z}+\Gamma_{I133}\dot{\tilde {D}}_{Iz},\notag \\
&\dot S_{I\theta r}=\mathcal{A}_{I2121}\frac{1}{r}\left(\frac{\partial u_r}{\partial \theta}-u_\theta\right)+\left(\mathcal{A}_{I1221}+p\right)\frac{\partial u_\theta}{\partial r}+\Gamma_{I122}\dot{\tilde {D}}_{I\theta},\notag \\
&\dot S_{I\theta z}=\mathcal{A}_{2323}\frac{1}{r}\frac{\partial u_z}{\partial \theta}+(\mathcal{A}_{I2332}+p)\frac{\partial u_\theta}{\partial z},\notag \\
&\dot S_{Izr}=\mathcal{A}_{I3131}\frac{\partial u_r}{\partial z}+(\mathcal{A}_{I1331}+p)\frac{\partial u_z}{\partial r}+\Gamma_{I133}\dot{\tilde {D}}_{Iz},\notag \\
&\dot S_{Iz\theta}=\mathcal{A}_{I3232}\frac{\partial u_\theta}{\partial z}+(\mathcal{A}_{I2332}+p)\frac{1}{r}\frac{\partial u_z}{\partial \theta},
\end{align}
and the components of the updated incremental electric field in \eqref{eq16-1}$_2$ as
\begin{align}\label{incremental-constitutive2}
&\dot E_{lIr}=-\frac{\partial\dot{\phi}}{\partial r}=\Gamma_{I111}\frac{\partial u_r}{\partial r}+\Gamma_{221}\frac{1}{r}\left(\frac{\partial u_\theta}{\partial \theta}+u_r\right)+\Gamma_{I331}\frac{\partial u_z}{\partial z}+\mathcal{K}_{I11}\dot{\tilde {D}}_{Ir},\notag\\
&\dot E_{lI\theta}=-\frac{1}{r}\frac{\partial\dot{\phi}}{\partial \theta}=\Gamma_{I122}\left[\frac{1}{r}\left(\frac{\partial u_r}{\partial \theta}-u_\theta\right)+\frac{\partial u_\theta}{\partial r}\right]+\mathcal{K}_{I22}\dot{\tilde {D}}_{I\theta},\notag\\
&\dot E_{lIz}=-\frac{\partial\dot{\phi}}{\partial z}=\Gamma_{I133}\left(\frac{\partial u_r}{\partial z}+\frac{\partial u_z}{\partial_r}\right)+\mathcal{K}_{I33}\dot{\tilde {D}}_{Iz}.
\end{align}

\newpage
\section*{Appendix B}
\label{B}
\appendix
\begin{align}
&\bm G_1=\left[ \begin{matrix}
 -1 & -m & -kr & 0\\
 m\left(1-\sigma_{rr}/\gamma_{12}\right) & (1-\sigma_{rr}/\gamma_{12}) & 0 & 0\\
 kr\left(1 - \sigma_{rr}/\gamma_{13}\right) & 0 & 0 & 0 \\
 \xi_1 & -\dfrac{m\sigma_{rr}}{\gamma_{12}}\dfrac{\Gamma_{I122}}{\mathcal{K}_{I22}} & 0 & 0
\end{matrix} \right],~
\\[12pt]
&\bm G_2=\left[\begin{matrix}
0 & 0 & 0 & 0 \\
0 & \dfrac{1}{\gamma_{12}} & 0 & -\dfrac{m}{\gamma_{12}}\dfrac{\Gamma_{I122}}{\mathcal{K}_{I22}} \\[12pt]
0 & 0 & \dfrac{1}{\gamma_{13}} & -\dfrac{kr}{\gamma_{13}}\dfrac{\Gamma_{I133}}{\mathcal{K}_{I33}}\\[12pt]
0 & \dfrac{m}{\gamma_{12}}\dfrac{\Gamma_{I122}}{\mathcal{K}_{I22}} & \dfrac{kr}{\gamma_{13}}\dfrac{\Gamma_{I133}}{\mathcal{K}_{I33}} & \xi_2
\end{matrix}\right], \notag 
\end{align}
\\[12pt]
{\footnotesize
\begin{align}
&\bm G_3=\left[\begin{matrix}
\kappa_{11} & \kappa_{12} & \kappa_{13} &-\left(\Gamma_{I111}-\Gamma_{I221}\right) \\
\kappa_{12} & \kappa_{22} & \kappa_{23} &-m\left(\Gamma_{I111}-\Gamma_{I221}\right) \\
\kappa_{13} & \kappa_{23} & \kappa_{33} &-kr\left(\Gamma_{I111}-\Gamma_{I331}\right)\\
\Gamma_{I111}-\Gamma_{I221} & m\left(\Gamma_{I111}-\Gamma_{I221}\right) & kr\left(\Gamma_{I111}-\Gamma_{I331}\right) & -\mathcal{K}_{I11}
\end{matrix}\right], \notag 
\end{align}
}
\\[12pt]
\begin{align}
&\bm G_4=\left[ \begin{matrix}
 1 & -\dfrac{m\left(\gamma_{12}-\sigma_{rr}\right)}{\gamma_{12}} & -\dfrac{kr\left(\gamma_{13}-\sigma_{rr}\right)}{\gamma_{13}} & \xi_1\\
 n &  -\dfrac{\gamma_{12}-\sigma_{rr}}{\gamma_{12}} & 0 & -\dfrac{n\sigma_{rr}}{\gamma_{12}}\dfrac{\Gamma_{I122}}{\mathcal{K}_{I22}}\\
 kr & 0 & 0 & 0 \\
 0 & 0 & 0 & 0
\end{matrix} \right],
\end{align}
in which 
\begin{align}\label{material-parameters}
&\gamma_{12}=\mathcal{A}_{I1212}-\dfrac{\Gamma_{I122}^2}{\mathcal{K}_{I22}}, \qquad \gamma_{21}=\mathcal{A}_{I2121}-\dfrac{\Gamma_{I122}^2}{\mathcal{K}_{I22}}, \quad \gamma_{23}=\mathcal{A}_{I2323}, \notag \\
&\gamma_{13}=\mathcal{A}_{I1313}-\dfrac{\Gamma_{I133}^2}{\mathcal{K}_{I33}}, \qquad \gamma_{31}=\mathcal{A}_{I3131}-\dfrac{\Gamma_{I133}^2}{\mathcal{K}_{I33}}, \quad \gamma_{32}=\mathcal{A}_{I3232},\notag \\
&\xi_1=-\left(\dfrac{\Gamma_{I122}}{\mathcal{K}_{I22}}\dfrac{m^2}{\gamma_{12}}+\dfrac{\Gamma_{I133}}{\mathcal{K}_{I33}}\dfrac{k^2r^2}{\gamma_{13}}\right)\sigma_{rr}, \notag \\
&\xi_2=-\left(\dfrac{m^2}{\mathcal{K}_{I22}}+\dfrac{\Gamma_{I122}^2}{\mathcal{K}^2_{I22}}\dfrac{m^2}{\gamma_{12}}+\dfrac{k^2r^2}{\mathcal{K}_{I33}}+\dfrac{\Gamma_{I133}^2}{\mathcal{K}_{I33}^2}\dfrac{k^2r^2}{\gamma_{13}}\right), \notag \\
& \beta_{12}=\dfrac{1}{2}\left(\mathcal{A}_{I1111}+\mathcal{A}_{I2222}-2\mathcal{A}_{I1122}-2\mathcal{A}_{I1221}+\dfrac{2\Gamma_{I122}^2}{\mathcal{K}_{I22}}\right), \notag \\
&\beta_{13}=\dfrac{1}{2}\left(\mathcal{A}_{I1111}+\mathcal{A}_{I3333}-2\mathcal{A}_{I1133}-2\mathcal{A}_{I1331}+\dfrac{2\Gamma_{I133}^2}{\mathcal{K}_{I33}}\right), \notag \\
&\kappa_{11}=2(\gamma_{12}-\sigma_{rr}+\beta_{12})+m^2\left[\gamma_{21}-\dfrac{\left(\gamma_{12}-\sigma_{rr}\right)^2}{\gamma_{12}}\right]+k^2r^2\left[\gamma_{31}-\dfrac{\left(\gamma_{13}-\sigma_{rr}\right)^2}{\gamma_{13}}\right], \notag \\
&\kappa_{12}=m \left(\gamma_{12}+\gamma_{21}+2\beta_{12}-\dfrac{\sigma_{rr}^2}{\gamma_{12}}\right), \notag \\
& \kappa_{13}=kr\left(\mathcal{A}_{I1111}+\mathcal{A}_{I2233}-\mathcal{A}_{I1122}-\mathcal{A}_{I1133}+p\right),\notag \\
&\kappa_{22}=2m^2(\gamma_{12}-\sigma_{rr}+\beta_{12})+\gamma_{21}-\dfrac{\left(\gamma_{12}-\sigma_{rr}\right)^2}{\gamma_{12}}+k^2r^2\gamma_{32}, \notag \\
&\kappa_{23}=m kr\left(\mathcal{A}_{I1111}+\mathcal{A}_{I2233}+\mathcal{A}_{I2332}-\mathcal{A}_{I1122}-\mathcal{A}_{I1133}+2p\right),\notag \\
&\kappa_{33}=2k^2r^2\left(\gamma_{13}-\sigma_{rr}+\beta_{13}\right)+m^2\gamma_{23}.
\end{align}
The detailed derivation of this Stroh formulation can be found in the paper by \cite{su2019finite}.

\end{document}